\documentclass[iop]{emulateapj}
\pdfoutput=1
\newcommand{\oii}{[\ion{O}{2}]}
\newcommand{\oiii}{[\ion{O}{3}]}
\newcommand{\hb}{H$\beta$} 
\newcommand{\ha}{H$\alpha$}
\newcommand{\lya}{Ly$\alpha$}

\newcommand{\mgii}{\ion{Mg}{2} $\lambda\lambda2796,2803$} 
\newcommand{\kms}{km s$^{-1}$} 
   
\newcommand{\fesclya}{$f_{esc}^{ \rm Ly \alpha}$}

\usepackage{amsmath} 

\begin{document} 
\title{A Close Relationship between \lya\ and M\lowercase{g} II in Green Pea Galaxies\altaffilmark{*}\altaffilmark{$\dagger$}} 
\author{Alaina Henry\altaffilmark{1,2}, Danielle A. Berg\altaffilmark{3},  Claudia Scarlata\altaffilmark{4}, Anne Verhamme\altaffilmark{5}, Dawn Erb\altaffilmark{3}} 

\altaffiltext{1}{Space Telescope Science Institute,  3700 San Martin Drive, Baltimore, MD,  21218, USA ahenry@stsci.edu} 
\altaffiltext{2}{Department of Physics \& Astronomy, Johns Hopkins University, Baltimore, MD 21218, USA}
\altaffiltext{3}{Center for Gravitation, Cosmology, and Astrophysics, Department of Physics, University of Wisconsin Milwaukee, 3135 Maryland Ave, Milwaukee, WI 53211, USA}
\altaffiltext{4}{Minnesota Institute for Astrophysics, School of Physics and Astronomy, University of Minnesota, 316 Church Str. SE, Minneapolis, MN 55455, USA} 
\altaffiltext{5}{Observatoire de Gen\`eve, Universit\'e de Gen\`eve, 51 Ch. des Maillettes, 1290 Versoix, Switzerland} 
\altaffiltext{*}{Based on observations made with the NASA/ESA Hubble Space Telescope, which is operated by the Association of Universities for Research in Astronomy, Inc., under NASA contract NAS 5-26555.  These observations are associated with program 12928, 11727, 13744, and 14201.}
\altaffiltext{$\dagger$}{Observations reported here were obtained at the MMT Observatory, a joint facility of the Smithsonian Institution and the University of Arizona.} 

\begin{abstract}
The \mgii\ doublet is often used to measure interstellar medium absorption in galaxies, thereby serving as a diagnostic for feedback and outflows. However,  the interpretation of \ion{Mg}{2} remains confusing, due to resonant trapping and re-emission of the photons, analogous to \lya.  Therefore,  in this paper, we present new MMT Blue Channel Spectrograph observations of \ion{Mg}{2} for a sample of 10 Green Pea galaxies at $z\sim0.2-0.3$, where \lya\  was previously observed with the Cosmic Origins Spectrograph on  {\it Hubble Space Telescope}. With strong, (mostly) double-peaked \lya\ profiles, these galaxies allow us to observe \ion{Mg}{2} in the limit of low \ion{H}{1} column density.  We find strong  \ion{Mg}{2}  emission and little-to-no absorption.  We use photoionization models to show that nebular \ion{Mg}{2} from \ion{H}{2} regions is non-negligible, and the ratios of \mgii/\oiii\ $\lambda$5007 vs.\  \oiii $\lambda$5007/\oii\ $\lambda$3727  form a tight sequence.  Using this relation, we predict intrinsic \ion{Mg}{2} flux, and show that \ion{Mg}{2} escape fractions range from 0 to 0.9.  We find that the \ion{Mg}{2} escape fraction correlates tightly with the \lya\ escape fraction, and the \ion{Mg}{2} line profiles show evidence for broader and more redshifted emission when the escape fractions are low.    These trends are expected if the  escape fractions and velocity profiles  of \lya\ and \ion{Mg}{2} are shaped by resonant scattering in the same low column density gas.  As a consequence of a close relation with \lya,  \ion{Mg}{2}  may serve as a useful diagnostic in the epoch of reionization, where \lya\ and Lyman continuum photons are not easily observed.   
\end{abstract} 

\section{Introduction} 
In the spectra of star-forming galaxies, interstellar and circumgalactic medium (ISM and CGM) absorption in the \mgii\ doublet has been used to study galactic outflows \citep{Weiner09, Rubin10, Rubin11, Coil, Erb12, Martin12, Kornei12, Tang, Zhu, Bordoloi16, Finley17}.    At $z\sim 1$,  alongside a complex of near ultraviolet (NUV)  \ion{Fe}{2} lines, the \ion{Mg}{2} lines are redshifted into the rest-frame optical where they are readily observed from the ground.    Consequently, as a diagnostic, \ion{Mg}{2} plays an important role for measuring the evolution of galactic outflows, feedback, and the CGM. 

Despite its potential usefulness, most studies acknowledge that, as a resonance feature, the \mgii\ doublet can be complicated to interpret.   
The lines are commonly observed with P-Cygni profiles, showing blueshifted absorption and redshifted emission.    This line shape is produced when the absorbed photons scatter in a moving envelope of gas.  The detection of P-Cygni emission implies that the pure absorption profile has been ``filled in'' by re-emission of absorbed photons.  Hence, absorption-line measurements of outflow velocity, column density, and line saturation can be incorrect \citep{Prochaska11, SP15}.   
Nevertheless, the \mgii\ doublet lines are strong, giving sensitivity to low column densities of gas.  
Modeling of the expanding medium generating the P-Cygni profile  can also constrain the 
geometry, velocity, and density of absorbing and emitting gas (\citealt{Prochaska11, SP15, Zhu}, Carr et al., submitted), thereby providing measurements of mass outflow rates.   Therefore, including \ion{Mg}{2} alongside other UV lines can be useful as we aim to leverage all available information to understand outflows and the CGM. However, these features {\it must} be interpreted carefully.  


\begin{figure*} 
 \includegraphics[scale=0.6, viewport=0 0 1000 350, clip]{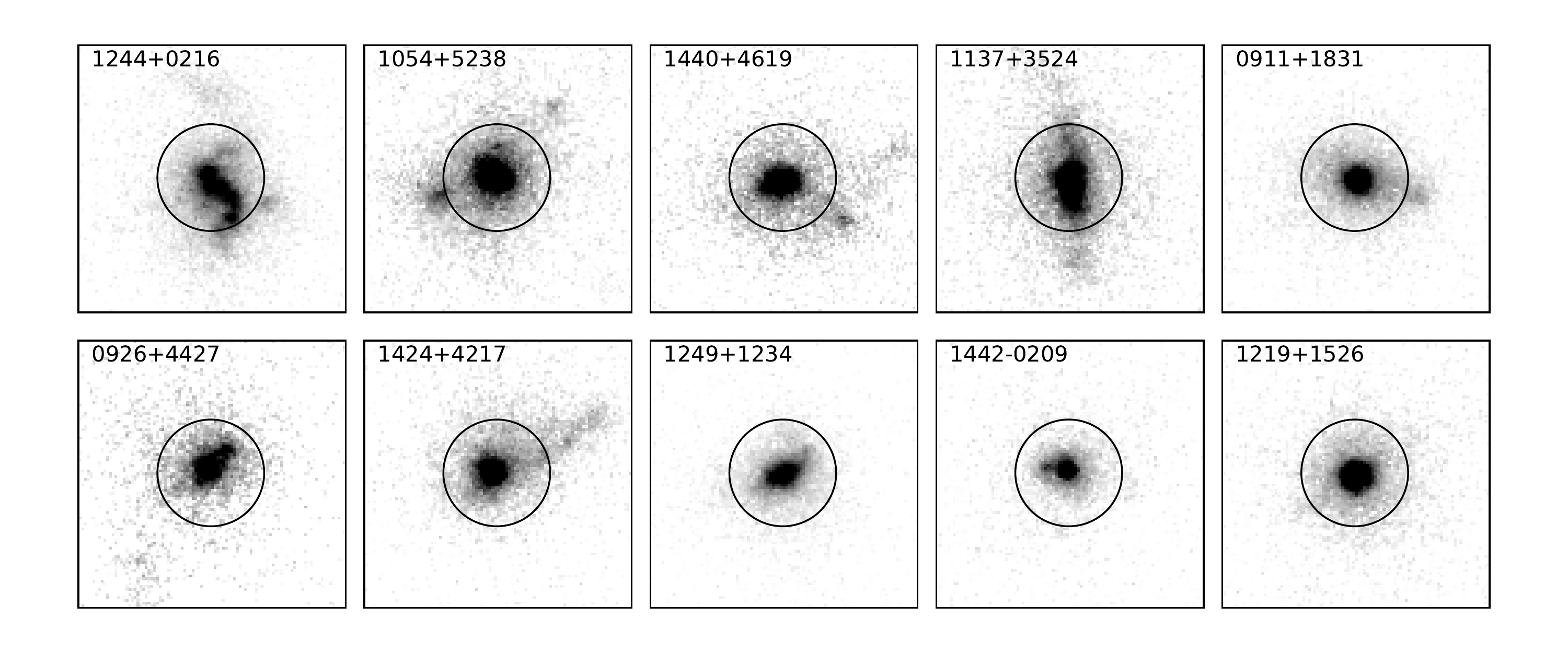}
\caption{Cutouts from the COS target acquisition images show the UV continuum light from the Green Pea galaxies in this sample.  Stamps are 2.5\arcsec\ on a side, similar to the diameter of the full COS aperture.   The circle marks the 1\arcsec\ diameter portion of the COS aperture that is unvignetted.  } 
\label{nuv_acq} 
\end{figure*}

The interpretation of \mgii\ is further complicated by  a possible contribution from nebular emission.  
Both \cite{Erb12} and \cite{Jaskot16} point out that photoionization models predict appreciable
 \mgii\ emission from \ion{H}{2} regions.  While this emission is erased by absorption in most cases, Erb et al. argue that it may account for the differences in the observed profiles of \ion{Mg}{2} and \ion{Fe}{2} 
 in galaxies at $z\sim2$.   
    Remarkably,  nebular  \ion{Mg}{2}  has been detected from some regions of Orion \citep{Boeshaar}, 
 and in some SDSS spectra of galaxies at $z>0.36$ \citep{Guseva}.  
   Recently,  \cite{Izotov18} report \ion{Mg}{2} emission from a  low-redshift Lyman Continuum leaker, and  \cite{Finley17} observe that \ion{Mg}{2} profiles transition from absorption to pure emission in $z\sim 1$ galaxies with $M \la 10^{9}$  $M_{\sun}$.  \ion{Mg}{2} emission is even seen in the supernova Refsdal host galaxy \citep{Karman}.

 Importantly, the production of nebular line photons, and subsequent resonant scattering in
 neutral/low ionization state gas is  identical to the \lya\ line.  In fact, some observations have suggested 
 that we may find similarities between these two features.    First,  studies of galaxies at $z\sim 1-2$ have reported that  
  \ion{Mg}{2}  emission is more common in galaxies with lower masses, higher specific star-formation rates, and blue colors \citep{Weiner09, Erb12, Kornei13, Finley17}. 
  These are the same properties that appear to favor \lya\ emission from galaxies  \citep{Gawiser07, Finkelstein07, Henry15}.  Second, spatially extended  \ion{Mg}{2}  has been detected in two 
intermediate redshift galaxies \citep{Rubin11, Martin13}, analogous to the ubiquitous \lya\ halos seen at both low and high redshifts \citep{Hayes14, Wisotzki, Leclercq}.    Nevertheless, in the only study that has compared \lya\ and  \ion{Mg}{2} to date, \cite{Rigby14} find no correlations for a sample of five galaxies.

\begin{deluxetable*} {cccccccccc}[!t]
\tablecolumns{9}
\tablecaption{Green Pea Sample} 
\tablehead{
\colhead{ID}  &  \colhead{RA}  & \colhead{DEC}  & \colhead{$z$\tablenotemark{a}} &\colhead{E(B-V)$_{MW}$\tablenotemark{b}}    & Exposure Time\tablenotemark{c} &  SDSS $u$ &  Slit Loss Correction\tablenotemark{d} & COS Gratings\tablenotemark{e} \\ 
  &   (J2000)  & (J2000) &   & (mag)  &    (seconds)       &  (mag)  & 
} 
\startdata
0911+1831 & 09 11 13.34 & 18 31 08.2 &    0.26224  &  0.025    & $4 \times 1800$  & 19.8 & 2.0  & G130M, G160M \\ 
0926+4427 & 09 26 00.44  &   44 27 36.5  &    0.18070  &  0.017 &   $4 \times 1800$ & 19.2 &2.4 & G130M, G160M  \\ 
1054+5238 & 10 53 30.80  &  52 37 52.9 &      0.25265  & 0.013   &   $3 \times 1800$ &19.0 & 2.0 & G130M, G160M  \\   
1137+3524  & 11 37 22.14 & 35 24 26.7 &     0.19440   & 0.016  &  $3 \times 900$ &19.1   &  1.5  & G130M, G160M  \\
1219+1526 & 12 19 03.98  & 15 26 08.5 &     0.19561   & 0.024  &  $4 \times 1800$ & 19.8 &  1.9 & G130M, G160M \\  
1244+0216 & 12 44 23.37  &   02 15 40.4  &     0.23942  & 0.021     &$3 \times 1800$  & 19.6   &  2.1  & G130M, G160M  \\
1249+1234 & 12 48 34.63  & 12 34 02.9 &     0.26340  & 0.025  & $4 \times 1800$ & 20.2  & 1.4   &  G130M, G160M \\
1424+4217  & 14 24 05.72 &  42 16 46.3 &     0.18480  & 0.009  &  $3 \times 2000$  & 19.3  &  1.6  & G130M \\ 
1440+4619  &  14 40 09.94  &  46 19 36.9 &  0.30076  & 0.012 &$ 3 \times  600$ & 19.8  & 1.8  & G160M  \\  
1442--0209\tablenotemark{f}  & 14 42 31.37  &  --02 09 52.8 & 0.29367 & 0.047  &  $3 \times 1800$ & 21.5   & 2.9   & G160M 
\enddata
\label{sample} 
\tablenotetext{a}{ The redshifts are determined from SDSS spectra, which have uncertainties of order a few \kms.}
\tablenotetext{b}{The Milky Way extinction values are from \cite{Schlafly}, acquired from the NASA Extragalactic Database.   We use the \cite{fm_unred} extinction curve to correct the UV spectra for foreground attenuation. } 
\tablenotetext{c}{An exposure time specified, for example, as $3 \times 1800$ implies that three 1800 second exposures were combined.} 
\tablenotetext{d}{The slit loss correction is estimated by calculating the SDSS $u-$band magnitude directly from each spectrum, and comparing this result to the imaged SDSS $u-$band magnitude.   } 
\tablenotetext{e}{The COS gratings used for comparison in this paper; low resolution gratings were not considered.}
\tablenotetext{f}{This Green Pea galaxy was drawn from the \cite{Izotov16} sample of Lyman Continuum leakers.} 
\end{deluxetable*}

For most redshifts, the wavelengths of \ion{Mg}{2} and \lya\ have made comparison between the lines difficult.  Consequently, 
more questions than answers remain about the interpretation of \ion{Mg}{2}, its similarities or differences with \lya\ and other ISM absorption lines, and its ability to serve as a diagnostic for outflows.    To address these issues, in this paper we present the first observations of \ion{Mg}{2} from $z\sim 0.2-0.3$ Green Pea galaxies.  With existing high-resolution, high signal-to-noise \lya\  and far ultraviolet (FUV) ISM absorption line spectra \citep{Henry15}, we are able to compare the \ion{Mg}{2} to other widely-used diagnostics.    This sample allows us to explore \ion{Mg}{2} in the limit of low \ion{H}{1} column density that has been inferred from \lya\ profiles  \citep{Henry15} and, in some cases, Lyman Continuum (LyC) leakage \citep{Izotov_nature, Izotov16, Izotov17, Izotov18}.  By focusing on this select set of rare, nearby galaxies, we can aim to better understand \ion{Mg}{2} under physical conditions that may be common at high-redshifts.  Ultimately, as we will demonstrate, \ion{Mg}{2} observations of these galaxies form a strong motivation for increased samples, covering a greater dynamic range of galaxy properties.

This paper is organized as follows.  In \S \ref{data} we present our sample, observations, and data reduction.    In \S \ref{char_mgii}  we show that the Green Pea galaxies have consistent strong \mgii\ emission and almost no absorption.  We present measurements of these emission lines, and predictions for intrinsic \ion{Mg}{2} strength using Cloudy photoionization models \citep{cloudy17}.    Using this intrinsic \ion{Mg}{2}  prediction, we calculate \ion{Mg}{2} escape fractions, analogous to \lya\ escape fractions.   In \S \ref{results}  we explore how the \ion{Mg}{2} escape fraction, equivalent width, and line velocity structure vary due to radiation transport effects, showing that \ion{Mg}{2} escape fractions form a tight sequence with \lya\ escape fractions. Finally, \S \ref{discussion} summarizes these results and discusses their implications, considering the utility of \ion{Mg}{2} as a diagnostic in the epoch of reionzation.  Throughout this paper, we adopt a convention where positive equivalent widths refer to emission, while negative equivalent widths indicate absorption.

\section{Sample, Observations, and Data} 
\label{data} 
Green Peas are a sample of $z\sim 0.2$ galaxies with extremely high-equivalent width emission lines \citep{Cardamone}, 
making them compelling analogs for high-redshift objects. 
We have selected ten such galaxies for observations in the \mgii\ doublet, using two main criteria.  First, we require that the objects have existing observations of \lya\ and UV continuum from the Cosmic Origins Spectrograph (COS) on the {\it Hubble Space Telescope}, with sufficient sensitivity to measure metal absorption lines.  Second, we chose objects with $z>0.18$, placing \mgii\ at observed wavelengths greater than 3300 \AA.   We chose eight of the ten galaxies from the Green Peas presented in \cite{Henry15}, adding one galaxy from Yang et al.\ (2017; HST GO 14201), and a LyC leaker from Izotov et al.\ (2016; HST GO 13744).    All of the supplemental COS observations use the high-resolution G130M and/or G160M gratings.  The sample is listed in Table \ref{sample}.    Figure \ref{nuv_acq} shows their COS NUV target acquisition images. 

We observed these ten galaxies using the Blue Channel Spectrograph \citep{bluechannel}  at the MMT on 28-29 March 2017.    
A 1\arcsec\ slit was used with the 832 lines mm$^{-1}$ grating at second order, achieving a spectral resolution of 90 km s$^{-1}$. The central wavelength of 3605 \AA\ provided wavelength coverage from 3150 to 4090 \AA\ in the observed frame.    Total exposure times were chosen to obtain an estimated continuum signal-to-noise (SNR) of ten per spectral resolution element, and were divided into three or four exposures per object.  Since atmospheric dispersion is significant at these wavelengths \citep{Fillippenko}, we ensured that the observations were carried out at the lowest possible airmasses (less than 1.25, except for 1442-0209 which reached 1.36).    We also reset the slit position angle for each exposure, choosing the parallactic angle midway through the exposure.  Likewise, because the Blue Channel Spectrograph does not compensate for flexure, arc lamp exposures were taken before the observation of each object.   Finally, we note that observing conditions were non-photometric, with light cirrus and variable seeing around 1\arcsec.

The spectra were processed using ISPEC2D \citep{MK06}, a long-slit spectroscopy data reduction package written in IDL,
and then extracted and calibrated using standard IRAF procedures.
A master bias frame was created from 10 zero-second exposures by discarding the highest and lowest value at each pixel and taking the median.
Master sky and dome flats were similarly constructed after normalizing the counts in the individual images.
Those calibration files were then used to bias-subtract, flat-field, and illumination-correct the raw data frames.
Due to faint continuum in individual exposures, the multiple sub-exposures of each target were median combined prior to extraction,
also eliminating cosmic rays in the process.
Extraction of the one-dimensional spectra was performed using a box-car aperture capturing 99\%\ of the light in the \ion{Mg}{2} feature
and a median sky subtraction along each column of the dispersion direction. 
For each object, the wavelength calibration was applied from the HeArHgCd comparison lamps taken at the same telescope pointing.
A sensitivity curve was derived from several standard stars observed during twilight,  and used to flux calibrate the individual 1D spectra. This spectrophotometric data reached 3200 \AA, so although our Blue Channel spectra reached 3150 \AA, we do not consider these wavelengths in any analysis.   Finally, the spectra were corrected to a heliocentric velocity reference frame, using IRAF's {\it rvcorrect}  procedure.

\begin{figure*}[!t]
\begin{center} 
 \includegraphics[scale=0.41, viewport= 50 10 1200 670, clip]{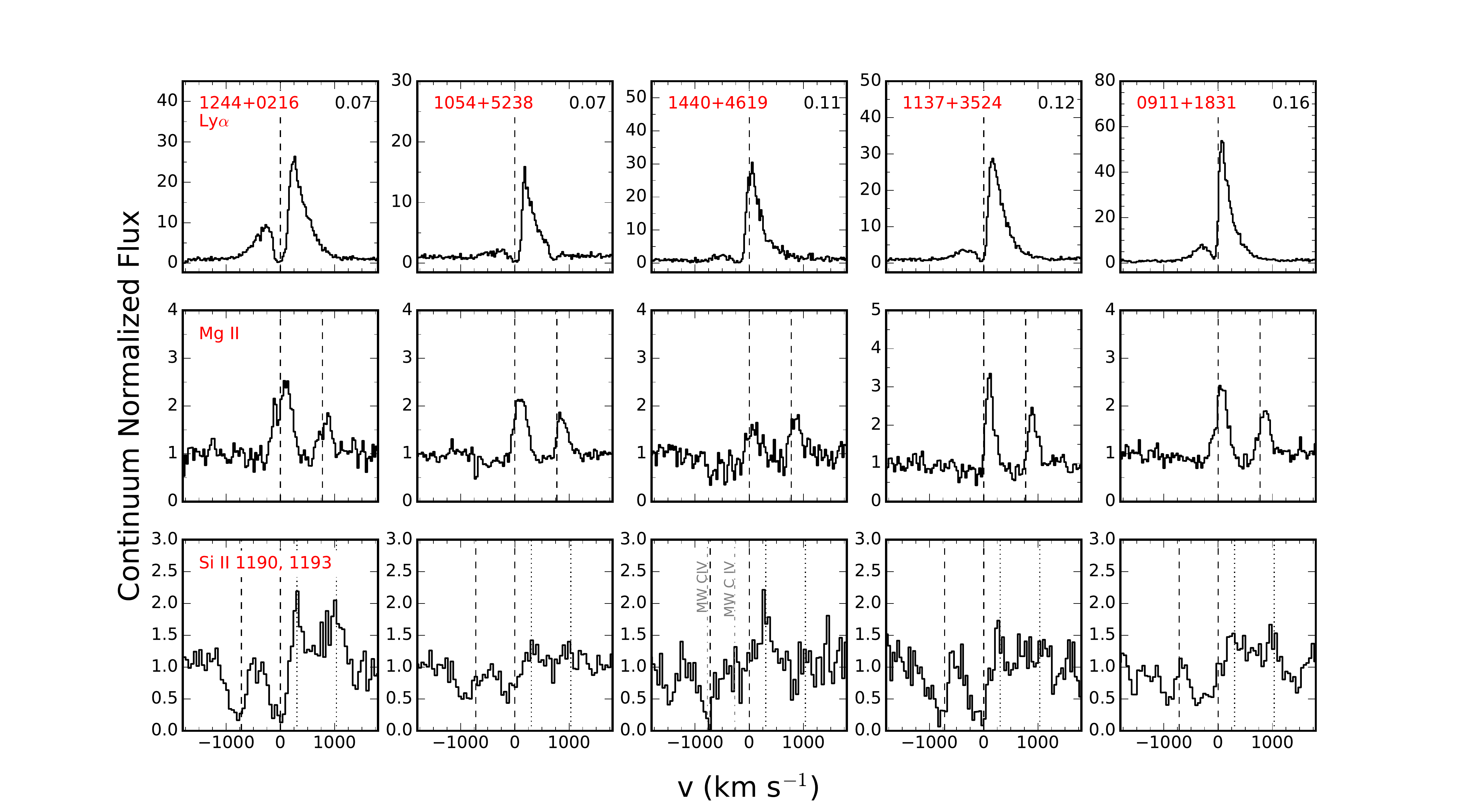} 
 \end{center}
\caption{The \mgii\ lines are compared to \lya\ and \ion{Si}{2}, plotted with respect to the systemic redshift at $v = 0$.  The top panels show \lya\ profiles observed with COS, taken from \cite{Henry15}, \cite{Yang17}, and \cite{Izotov16}.   The middle row shows the \mgii\ doublet, with the velocity scale set relative to the 2796 \AA\ line.  The bottom row shows the \ion{Si}{2} lines, observed in the same COS spectra as the \lya.   These spectra are ordered with \fesclya\ increasing from left to right, with measured values given in the upper right hand corner of each \lya\ panel.   These five galaxies represent the half of the sample with the lowest \fesclya.
Dashed vertical lines mark zero velocity for the emission or absorption features.   Only some spectra show detectable \ion{Si}{2} absorption, and some also show evidence of fluorescent \ion{Si}{2}* $\lambda \lambda$ 1194, 1197 emission (dotted lines).   Milky Way contamination from \ion{C}{4} and \ion{Co}{2} is noted by grey dot-dashed lines in two of the \ion{Si}{2} spectra. }
 \label{mgii_fig1} 
 \end{figure*} 
Since we aim to measure Doppler shifts for the \ion{Mg}{2} lines, we tested the accuracy of our wavelength calibration.  For this check, we compared our data to the night sky spectrum from \citealt{uves}, taken with the Ultraviolet and Visible Echelle Spectrograph (UVES) on the Very Large Telescope (VLT).   This spectrum has an average wavelength accuracy of 17 m\AA\ (1.5 \kms\ at 3400 \AA).  First, we downgraded this high resolution spectrum to 90 \kms, and rebinned it to match the Blue Channel data.  Then we cross-correlated the spectra in 200 \AA\ segments to derive any residual velocity shifts between the UVES and Blue Channel data.   We find that, for some wavelengths, the VLT and MMT sky spectra show substantial time, airmass and/or site-dependent variations that limit the utility of this method.  Nevertheless, at wavelengths where the spectra appear to have a similar spectral-morphology, this test indicates that the calibration in our data is accurate to 20 \kms.

In order to compare the MMT spectra with SDSS and COS observations, we calculate slit-loss corrections to place the Blue Channel spectra on an absolute flux scale.  We convolve each spectrum with the SDSS $u-$band throughput   ($\lambda_{center} \sim 3600$ \AA, FWHM $\sim 500$ \AA), 
calculating a synthetic $u-$band magnitude for the Blue Channel data.  Comparing to the SDSS photometry, we find that a factor of around two is required to correct for slit-losses and non-photometric conditions.  The SDSS $u-$band magnitudes and slit-loss corrections are given in Table \ref{sample}, and are applied to the \ion{Mg}{2} fluxes in \S \ref{char_mgii}.  Although we do not detect \ion{Mg}{2} emission extending beyond the continuum, we acknowledge that these corrections could still be lower limits for the lines, if extended, low surface brightness  line emission is missed by our observations.

Finally, we combine these observations with COS spectroscopy covering \lya\ and FUV absorption lines.  We described the data in detail in \cite{Henry15}.   In brief, we estimate  that the spectral resolution is 20-40 \kms\ for compact continuum emission, degraded only somewhat from the optimal point source spectral resolution.    Measurements of the Milky Way ISM absorption lines and CIII 1175.5 \AA\ photospheric absorption in the Green Peas implies a redshift precision around 30 \kms.    We rebin the spectra in bins of 20 pixels for continuum and absorption lines (around 45 \kms) and 10 pixels for \lya.   Noise in the binned spectra is estimated by summing the photon counts in each bin and calculating the Poisson noise according to \cite{Gehrels}.  The Green Peas 1440+4619 and 1442-0209, which were not originally in the \cite{Henry15} sample, have been treated in the same manner.    We do not correct the COS spectra for aperture losses, which, for \lya,  are mostly unknown, but significant when measured \citep{Henry15}.   As a result, the \lya\ measurements give a sense of the flux that escapes from the central few kpc of these compact galaxies.

\section{Analysis} 
\label{char_mgii} 
The \ion{Mg}{2}, \lya, and \ion{Si}{2} spectra are shown in Figures \ref{mgii_fig1} and \ref{mgii_fig2}.    The most striking observation in these figures is the strong \ion{Mg}{2} emission and weak (or absent) absorption, similar to the \ion{Mg}{2} emitters reported by \cite{Erb12} and \cite{Finley17}.    
This finding strongly suggests that \ion{Mg}{2} originates as an emission line, which is not erased by absorption in these galaxies. 
 However, we still see effects of scattering on lines:  in each galaxy, the \ion{Mg}{2} lines are systematically redshifted.   Moreover, the FWHM of the lines range from 100 to 300 \kms,  in  some cases, broader than we observe in the optical nebular lines.   In contrast, however, we see no evidence that \ion{Mg}{2} is extended beyond the continuum (e.g. \citealt{Rubin11, Martin13}). Remarkably, the \ion{Mg}{2} spectra bear no resemblance to the \ion{Si}{2} lines shown in the bottom rows of  Figures   \ref{mgii_fig1} and \ref{mgii_fig2}.   We will return to this puzzling observation in \S \ref{results}. 

\begin{figure*}[!t]
\begin{center} 
\includegraphics[scale=0.41, viewport=50 10 1200 670, clip]{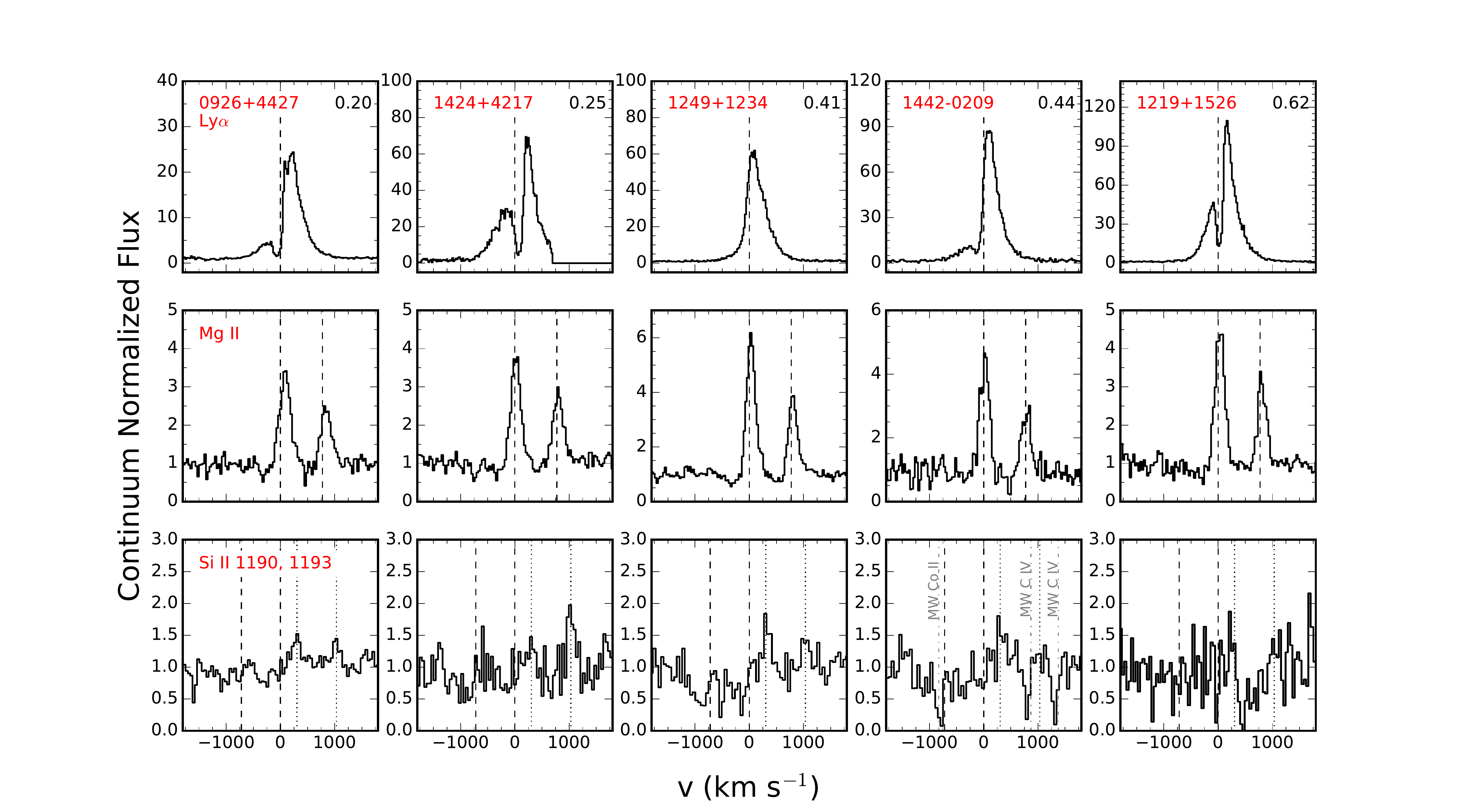} 
\end{center}
\caption{Continued from Figure \ref{mgii_fig1}.  These five galaxies represent the half of the sample with the highest  $f_{esc}^{Ly\alpha}$. } 
\label{mgii_fig2}
\end{figure*}

\subsection{Measurements} 
To further understand the unusual \ion{Mg}{2} emission, we measure fluxes, equivalent widths, peak velocities, and FWHM from these lines.  Each spectrum is normalized around the \ion{Mg}{2} lines using a linear fit to continuum regions spanning several thousand \kms\ around the systemic velocity.   We find
that the statistical uncertainty on the continuum normalization is a few percent or better, although visual inspection of the data suggests an approximately 5\% systematic uncertainty.  Next,  we divide the spectrum into two regions corresponding to each line, using the midpoint between the two lines, $\lambda_{rest} = 2799.1$ \AA.   Bluer wavelengths are assigned to the 2796 \AA\ line while redder wavelengths are assigned to the 2803 \AA\ line.  Then, we calculate the equivalent width weighted velocity of each line, $v_{peak}$, as a nonparametric means of centroiding the emission (see \citealt{Henry15}).   Since some lines show minimal absorption blueward of the 2796 \AA\ line, we measure a  blueshifted velocity where the absorption meets the continuum. For the 2796 \AA\ line, the fluxes and equivalent widths are integrated to this maximal velocity.  In the case of 1440+4619, the emission is mostly erased by this absorption.  
 Finally, errors are calculated through a Monte Carlo simulation where we perturb the spectrum according to its errors, repeat the measurements, and calculate the RMS.

The \ion{Mg}{2} measurements are likely impacted by stellar \ion{Mg}{2} absorption.   \cite{MB09} highlight how this absorption varies with the characteristic age of the stellar population, showing that weaker absorption corresponds to younger stellar populations.   In order to quantify this absorption for the present sample, we 
measured the absorption equivalent width in the Binary Population and Spectral Synthesis (BPASS v2.0; \citealt{bpass}) models that we use for photoionization modeling in \S \ref{cloudy_sec} (see below).     At the resolution of these models, the \ion{Mg}{2} lines are blended.  Nevertheless, we find a minimum  $W_{stellar} = 0.2$ at ages of a few Myr (for both bursts and continuous star-forming models).   An upper limit on  $W_{stellar}$ can be inferred by considering the maximum plausible ages that can still
produce  $W_{H \alpha} \gtrsim 300$ \AA, typical for Green Peas.   This requirement implies that the ages can be at most a few hundred Myrs for continuous star-forming models, which corresponds to  $W_{stellar} < 1.0$ in the BPASS models considered here.   Therefore, we adopt   $W_{stellar} = 0.6$; since the \ion{Mg}{2} lines are unresolved in the stellar models, we apply a 0.3 \AA\ correction to the equivalent width of each \ion{Mg}{2} line, and scale the measured fluxes accordingly. 
The \ion{Mg}{2} measurements are presented in Table \ref{mgii_table}.


\begin{figure*}
\centering 
\includegraphics[scale=0.8, viewport=0 0 600 500, clip]{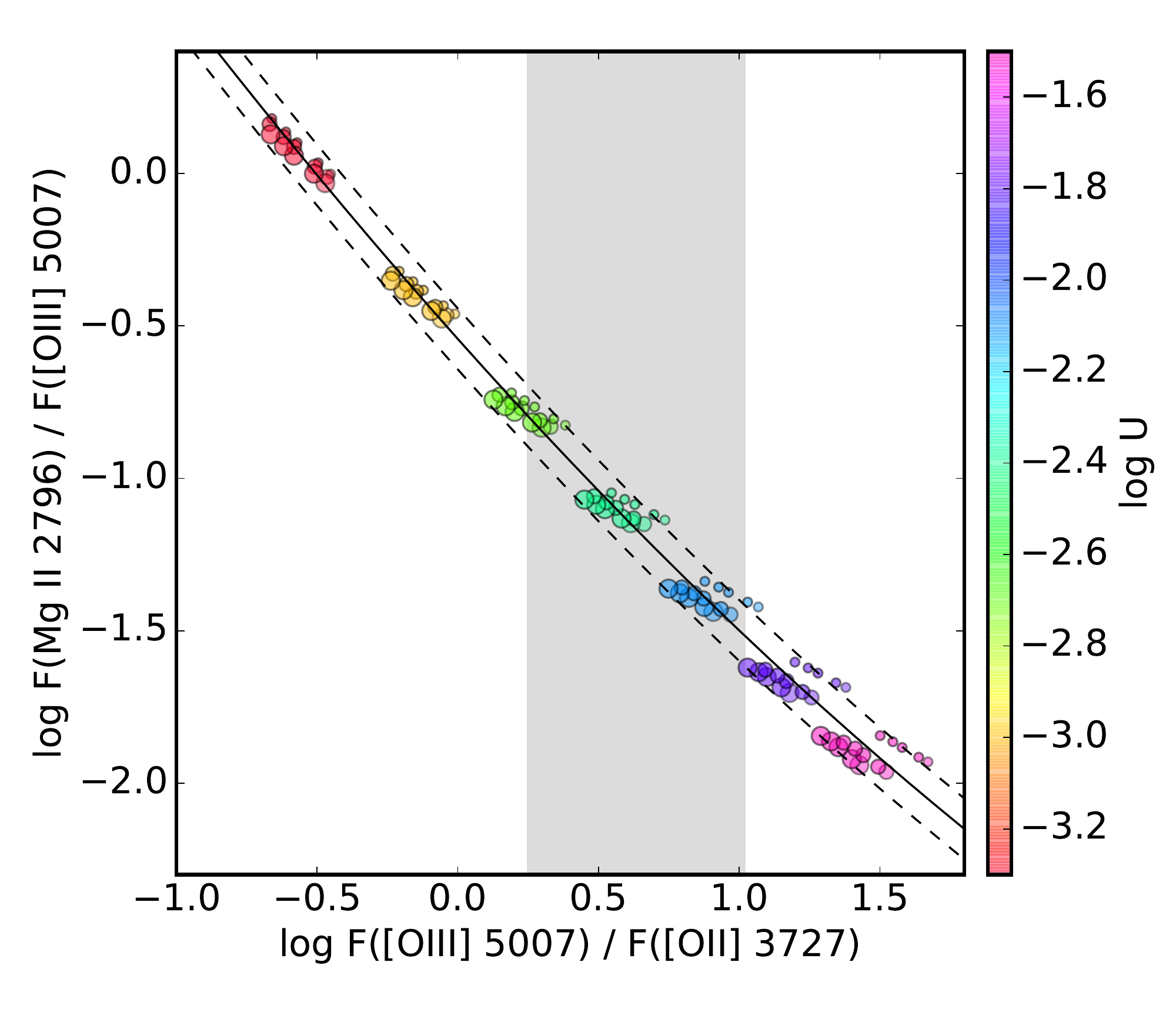}
\caption{Cloudy photoionization models show that \ion{Mg}{2}/[\ion{O}{3}] forms a tight sequence with [\ion{O}{3}]/[\ion{O}{2}].  For simplicity, and since all BPASS and Starburst99 models fall along the same sequence, we show only a subset of the BPASS (version 2.0) models,  using  the continuous star-formation histories, a 100  M$_{\sun}$ upper mass cutoff, and $n_e = 10^{2.4}$ cm$^{-3}$. The grey shaded region shows the range of  [\ion{O}{3}]/[\ion{O}{2}] values covered by our sample.   The different colored points highlight the variations in ionization parameter, $U$, and the point sizes are proportional to the gas-phase metallicity (Z/Z$_{\sun}$ =  0.1, 0.25, 0.4).   Each grid point (log $U$ and gas-phase metallicity) is plotted for five stellar metallicities (Z = 0.001, 0.002, 0.004, 0.006, and 0.008).  In these models, the harder ionizing spectrum at lower stellar metallicities moves the line ratios downwards along the sequence.   The solid line indicates the polynomial fits to the BPASS models (Equations  \ref{mg2_fit_1} and \ref{mg2_fit_2}), while the dashed lines mark $\pm 0.1$ dex  }
\label{cloudy_bpass} 
\end{figure*}

For the COS spectra, we compile several measurements from \cite{Henry15}, repeating that analysis for 1440+4619 and 1442-0209.  First, we include the \lya\ escape fractions, \fesclya\  = F(\lya) /  (8.7 $\times$ F(H$\alpha$)$_{corr}$),  where F(\ha)$_{corr}$ is corrected for dust (according to \citealt{Calzetti}). For this calculation, we use an intrinsic \lya/\ha\ ratio of 8.7, since these Green Pea galaxies have elevated electron densities ($n_e = 100 - 700$ cm$^{-3}$), which increases  \lya/\ha\ above the Case-B low density limit of 8.1 \citep{DS03,Henry15}.   The \lya, \ion{Mg}{2}, and \ion{Si}{2} spectra in Figures \ref{mgii_fig1} and \ref{mgii_fig2} are shown in order of increasing \fesclya.   Second, in addition to \lya\ escape fractions, we include kinematic measures for the \lya\ lines.   We measure the peak velocities,  $v_{blue}^{peak} $ and $v_{red}^{peak}$, since peak separation is sensitive to the \ion{H}{1} column density \citep{Verhamme15}, and has been shown to correlate with \fesclya\ \citep{Henry15}.   Finally, velocities and equivalent widths of FUV absorption can also be compared to \ion{Mg}{2}.    For  1440+4619 and 1442-0209, we either do not convincingly detect the low-ionization  (LIS) metal absorption lines (\ion{Si}{2} and \ion{C}{2})  that trace \ion{H}{1}, or we find that the spectra are contaminated by foreground Milky Way absorption.  Therefore, we either set upper limits on the equivalent widths at $W_{\rm LIS} \ga -0.5$ \AA, or exclude the lines.   For the remainder, we take the measurements from \cite{Henry15}, and do not reproduce that tabular data here.  Since these lines are often weak in Green Peas and similar galaxies \citep{Erb10, Henry15, Chisholm17},  we mostly focus on \lya\ in this paper, but we do consider whether \ion{Mg}{2} varies with $W_{\rm LIS}$ in \S \ref{results}.  To conclude, the \ion{Mg}{2} and \lya\ measurements are reported in Tables \ref{lya_table} and \ref{mgii_table}.

\subsection{Calculating \ion{Mg}{2} Escape Fractions with Photoionization Models} 
\label{cloudy_sec} 
In order to understand how the observed \ion{Mg}{2} varies due to scattering and absorption in
the ISM and CGM, we must first account for the intrinsic variations in the \ion{Mg}{2} produced in \ion{H}{2} regions.  Therefore, we aim to predict the intrinsic \ion{Mg}{2} fluxes, using  photoionization models constrained by the physical properties determined from the SDSS spectra (i.e., oxygen abundance and electron density, $n_e$).   Fortunately, for the Green Peas in our sample,  significant detections of the temperature-sensitive \oiii\ $\lambda$4363 auroral line allow accurate determinations of the oxygen abundance using the ``direct'' method\footnote{While there has been a long-standing debate over the accuracy of the ``direct'' method, recent studies have shown remarkable agreement between stellar and ``direct'' method nebular abundances in typical nearby galaxies \citep{Davies17, Zahid17}.  Therefore, we judge the gas-phase metallicity calibration as accurate to around 0.2 dex. }.
We follow the procedure described in \cite{Berg15}\footnote{Specifically, we use the IDL-based IMPRO routines (\url{https://github.com/moustakas/impro}), along with the atomic data in \cite{Berg15}.} to determine electron densities and direct method abundances.   For this calculation, we correct the emission lines for internal dust attenuation using the reddening law of \cite{Calzetti}, and also use the theoretical  \cite{Garnett92} relationship to estimate the [\ion{O}{2}] electron temperature from the measured [\ion{O}{3}] temperature.  The electron densities were measured from the  [\ion{S}{2}] $\lambda \lambda$6716, 6731 doublet, and ranged from 100-700 cm$^{-3}$.  The metallicities are between 12 + log(O/H) = 7.9 to 8.2 ($Z_{gas} = 0.16-0.32 Z_{\sun}$,  where $Z_{\sun}$ corresponds to 12 + log(O/H) = 8.69; \citealt{Grevesse}), and are listed in Table \ref{lya_table}. 
 
For photoionization modeling, we use Cloudy version 17 \citep{cloudy17}, considering a range of ionizing spectra with sub-solar metallicities.   
First, we include fourteen models from Starburst99 \citep{Leitherer99,Leitherer14}.  In brief, these are: the Padova models, with and without AGB stars, at  $Z_{stars} = 0.0004, 0.004, 0.008$ { ($Z = 0.029, 0.29$, and 0.57 $Z_{\sun}$, where $Z_{\sun}$ = 0.014)}; the Geneva models with standard and high mass loss rates at $Z_{stars} = 0.001, 0.004, 0.008$;  and the Geneva models with and without rotation at $Z_{stars}=0.002$ ({$Z = 0.14 Z_{\sun}$}).   In all Starburst99 models, we use the \cite{Kroupa} stellar initial mass function (IMF). Second, we consider the Binary Population and Spectral Synthesis models (BPASS v2.0 {\it and } v2.1; \citealt{bpass, bpass2p1}), which include the effects of binary evolution and produce harder ionizing spectra than Starburst99.  For these models, we include metallicities of  $Z_{stars} = 0.001, 0.002, 0.004, 0.008$ ($Z = 0.07, 0.14, 0.29$ and 0.57 $Z_{\sun}$) and use the fiducial IMF, which has a slope of $-1.30$ between 0.1 and 0.5$M_{\sun}$ and $-2.35$ above 0.5$M_{\sun}$.  For BPASS v2.1, we also include the $Z_{stars} = 10^{-4}$ and $10^{-5}$ models.  For all BPASS models, we consider the IMFs both with 100 and 300M$_{\sun}$.   Finally,  we select both constant star-forming models and bursts aged 1, 3, and 10 Myrs for both BPASS and Starburst99, except for BPASS v2.1, where only the bursts are available. 

Following \cite{Steidel16},  the photoionization models were run using a  plane-parallel geometry.  The ionization parameter, $U$, was varied between $ -3.3  < {\rm log} ~ U < -1.5$ in 0.3 dex increments.  Models were calculated for three gas-phase metallicities, allowing for systematic uncertainties by covering a slightly broader range than our measurements: $Z_{gas} =0.1, 0.25,~ {\rm and}~ 0.4 ~Z_{\sun}$.  Likewise, we calculated  models over the range of densities measured for our sample: $n_{e} = 10^{2.0}, 10^{2.4} , ~{\rm and} ~10^{2.8}$ cm$^{-3}$.   Finally, we chose the Orion dust grains,  and a solar relative abundance pattern \citep{Grevesse}.  Here, we  do not require the stellar and gas-phase metallicities to match,  as non-solar $\alpha$/Fe abundances can give the appearance of decoupled stellar and nebular metallicities \citep{Steidel16}. 

We explored a number of line diagnostic diagrams in order to predict \ion{Mg}{2} from the optical 
spectrum of the Green Peas.  This exercise showed that \ion{Mg}{2}/[\ion{O}{3}] $\lambda$5007 forms a tight sequence with the ratio [\ion{O}{3}] $\lambda$5007/[\ion{O}{2}] $\lambda$3727, as illustrated in Figure \ref{cloudy_bpass}.   Here, we show \ion{Mg}{2} $\lambda 2796$ /[\ion{O}{3}] $\lambda$5007.  The same plot with \ion{Mg}{2} $\lambda 2803$ /[\ion{O}{3}] $\lambda$5007 appears similar, with only a factor of two difference set by the intrinsic \ion{Mg}{2} $\lambda 2796$ / \ion{Mg}{2} $\lambda 2803$ ratio.  
We find that this sequence does not depend  on the stellar model, as harder ionizing spectra move the line ratios down and to the right {\it along} the sequence.  This effect can be seen in Figure \ref{cloudy_bpass},  as each set of colored points (a single ionization parameter) is shown in a cluster of 5 ``rows'', each corresponding to a BPASS v2.0 model with a different stellar metallicity.   Likewise, gas-phase metallicity moves the line ratios roughly orthogonally to the sequence.  Hence, metallicity plays a key role in setting the scatter in Figure \ref{cloudy_bpass}.   We have conservatively considered gas-phase metallicities 0.1 dex lower and higher than we have measured, as we will use this scatter below to estimate the uncertainty on the predicted intrinsic \ion{Mg}{2} flux.

\begin{figure*} 
\centering
\includegraphics[scale=0.49, viewport= 20 0 1200 430, clip]{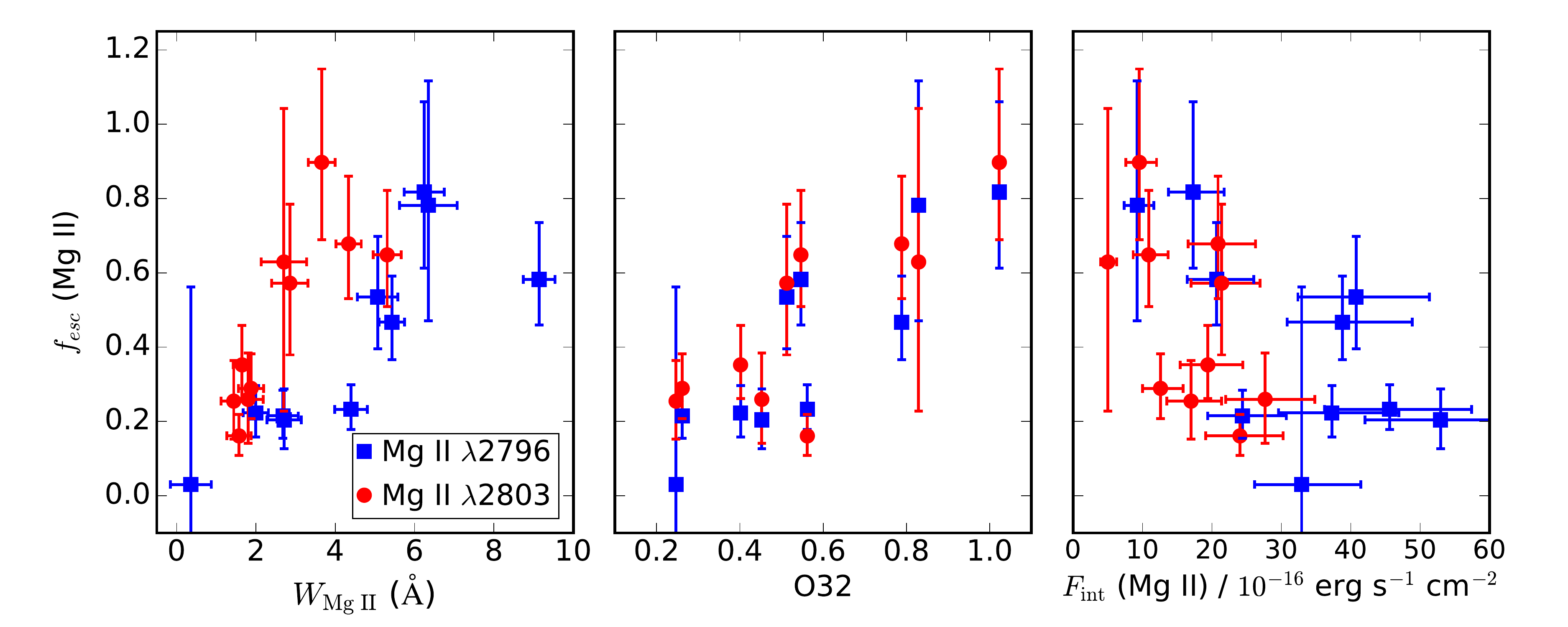} 
\caption{{\it Left--} The escape fraction of the \ion{Mg}{2} lines correlates with their equivalent width, indicating that stronger emission results from greater escape rather than intrinsically strong emission.  {\it Center--} Higher \ion{Mg}{2} escape fraction is also associated with higher O32 ($\equiv$ \oiii\ $\lambda 5007$/\oii\ $\lambda 3727$) ratios, leading to weaker \ion{Mg}{2}/\oiii\ ratios (following Figure \ref{cloudy_bpass}  and Equations \ref{mg2_fit_1} and \ref{mg2_fit_2}).  {\it Right--} These weaker 
\ion{Mg}{2}/\oiii\ ratios translate to fainter  intrinsic \ion{Mg}{2} flux in the sources with the highest \ion{Mg}{2} escape fractions. 
 In all panels, the uncertainties on the \ion{Mg}{2} escape fraction include a 0.1 dex uncertainty on the predicted intrinsic \ion{Mg}{2} flux, added in quadrature to the measurement error.}
\label{mgii_ew_fesc_2}
\end{figure*}

Figure  \ref{cloudy_bpass}  shows only continuous star-forming BPASS models with $n_{e} = 10^{2.4}$.   Nevertheless, for the ionizing spectra with sub-solar metallicities that we consider here, all reasonable Starburst99 and BPASS models fall along the same sequence, regardless of star-formation history or IMF upper limit.  Still, we verified that this sequence is robust to several other changes in the models.  First, changing $n_{e}$ shifted the line ratios by only a negligible amount.   
{Second, we  considered the effects of reduced dust depletion, as \cite{deCia} show that dust-to-metal ratios may be lower at lower metallicity.  We ran a subset of Cloudy models with no grains, finding that \ion{Mg}{2}/\oiii\ ratios would be higher by around 0.1-0.2 dex over the O32 range covered by our sample.  However, since this limiting case is extreme, we infer that uncertainties in dust depletion probably lead to line ratios that fall within the 0.1 dex scatter in Figure \ref{cloudy_bpass}.    
Third, we explored how an active galactic nucleus  (AGN) may impact the inference of intrinsic \ion{Mg}{2}.    We ran Cloudy models, extending to gas-phase metallicity of $Z_{gas} = 2.0 Z_{\sun}$, and using Cloudy's default AGN spectral template: $T = 1.5 \times 10^5$ K for the  ``Big Bump'' component,  an X-ray to UV ratio, $\alpha_{ox} = -1.4$, and the low-energy slope of $\alpha_{uv}  = -0.5$.  We find that the O32 range covered by our sample is reached at super solar gas-phase metallicities, with \ion{Mg}{2}/\oiii\ ratios 0.2-0.3 dex lower than the sequence in Figure \ref{cloudy_bpass}.  Therefore, this AGN model would predict weaker intrinsic \ion{Mg}{2} emission by a factor of 1.5-2, possibly even weaker than the observed \ion{Mg}{2} fluxes.  We note, however, that the Green Peas show no evidence for the presence of an AGN.  }    Finally,  we tested some ionizing spectra with super-solar metallicities, and found a number of model grid-points may fall below the tight sequence in Figure  \ref{cloudy_bpass}.  Likewise, the 10 Myr aged burst, with BPASS v2.1 spectra does not produce realistic line ratios for the Green Peas  (e.g. \oiii/\hb $<$ 1).  Hence, we exclude these models from consideration.  We conclude that the sequence in Figure \ref{cloudy_bpass} should be robust for our sample.

Figure \ref{cloudy_bpass} demonstrates that, with [\ion{O}{3}] $\lambda$5007/[\ion{O}{2}] $\lambda$3727 alone, we can predict the intrinsic  \ion{Mg}{2}/[\ion{O}{3}] $\lambda$5007 to within 0.1 dex for these Green Peas. This figure also shows the polynomial fit to our full set of BPASS v2.0 model grids, for the IMF with a 100 M$_{\sun}$ upper mass cutoff.  These relations are: 
\begin{equation} 
{\rm R_{2796}}   = 0.079 \times  {\rm O32}^2   - 1.04\times {\rm O32}  - 0.54 
\label{mg2_fit_1} 
\end{equation} 
and 
\begin{equation} 
{\rm R_{2803}}   = 0.098 \times  {\rm O32}^2   \\
  - 1.02\times {\rm O32}  - 0.84, 
\label{mg2_fit_2} 
\end{equation}  
where  
\begin{equation}   
\begin{split} 
{\rm R_{2796} = log ~ (Mg~II ~\lambda 2796 /[OIII] ~\lambda 5007)},  \\
{\rm R_{2803} = log ~ (Mg~II ~\lambda 2803 / [OIII] ~\lambda 5007)},  \\ 
{\rm and}  \\ 
{\rm O32} =  {\rm log}  ~([{\rm O III}]~ \lambda 5007 / [{\rm O II}]~ \lambda 3727). 
\end{split} 
\end{equation} 
We reiterate that these relations are valid over the gas-phase metallicities that we consider here: 12+log(O/H) = 7.8 - 8.3 ($Z = 0.1 - 0.4 Z_{\sun}$).  Higher gas-phase metallicities tend to move points down and to the left in Figure  \ref{cloudy_bpass}, increasing the scatter to a value more like 0.2 dex.  In this case, deriving a unique best-fitting Cloudy model for each galaxy may be a better strategy than using the relations given here.

Finally, given the intrinsic  \ion{Mg}{2}/[\ion{O}{3}] $\lambda$5007 ratios, we can calculate the fraction of \ion{Mg}{2} photons that escape the galaxies.    We use the  \oiii\ and \oii\ lines  (corrected for dust using \citealt{Calzetti})  to predict the intrinsic \ion{Mg}{2} for each line of the doublet, in each galaxy.  Then, we calculate the escape fractions by comparing observed and intrinsic line fluxes: 
\begin{equation} 
\begin{split} 
f_{esc}^{2796} = F^{obs} ({\rm Mg~II}~ 2796) / F^{int} ({\rm Mg~II}~ 2796)  \\ 
{\rm and} \\ 
f_{esc}^{2803} = F^{obs} ({\rm Mg~II}~ 2803) / F^{int} ({\rm Mg~II}~ 2803). 
\end{split}
 \end{equation} 
In this calculation, the observed  \ion{Mg}{2} fluxes are corrected for  slit-losses as described in \S \ref{data}.  We also correct for Milky Way foreground extinction, using the \cite{fm_unred} attenuation curve and the extinction values from (\citealt{Schlafly}; E(B-V)$_{MW}$ listed in Table \ref{sample}).   Additionally, we note that the uncertainty on \ion{Mg}{2} escape fraction includes the statistical error on the line fluxes and the 0.1 dex uncertainty on the intrinsic \ion{Mg}{2} prediction.  We do not, however, include an error to account for systematic uncertainties in the internal extinction law, as this error is ill-defined.   Although we chose \cite{Calzetti}, we note that other attenuation curves (e.g.\ SMC, LMC; \citealt{Gordon}, Milky Way; \citealt{ccm, fm_unred}) predict smaller absolute dust corrections, leading to lower intrinsic \oiii\ fluxes and up to 25\% higher \ion{Mg}{2}  escape fractions.  The \ion{Mg}{2} escape fractions are given in Table \ref{mgii_table}.

Overall, our calculation of \ion{Mg}{2} escape fractions is analogous to the calculation of \lya\ escape fractions in its treatment of dust. The Cloudy models presented in Figure \ref{cloudy_bpass}, and the relations in Equation \ref{mg2_fit_1} and \ref{mg2_fit_2}  use intrinsic line fluxes, rather than dust attenuated.  Likewise, the observed \lya\ and \ion{Mg}{2} line fluxes are not corrected for internal dust extinction, since a robust correction is difficult to discern when photons are resonantly scattered.  Therefore, like \lya, an escape fraction less than unity implies that photons are absorbed by dust, or scattered outside the aperture.    

We judge these estimates to be accurate in a relative sense, but acknowledge that photoionization modeling of the intrinsic line fluxes is subject to systematic uncertainties.  Different assumptions about relative abundance ratios, dust depletion patterns, and shocks may predict systematically different intrinsic \ion{Mg}{2} emission.  Hence, we urge caution, and focus on qualitative trends going forward.

\begin{figure} 
\centering
\includegraphics[scale=0.7, viewport=0 0 325 575, clip]{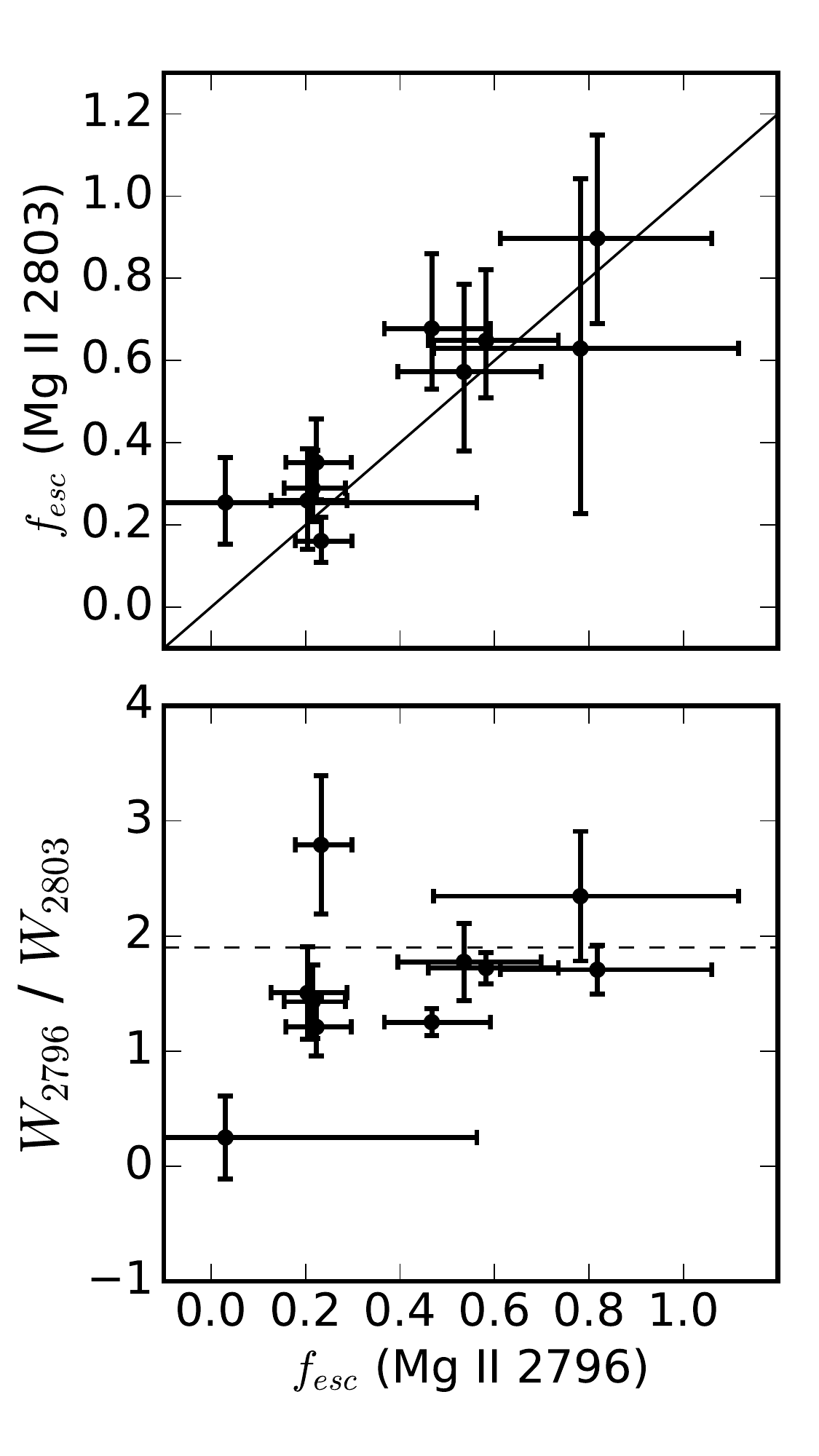}
\caption{{\it Top:} The  \ion{Mg}{2} $\lambda$2796  and $\lambda 2803$ escape fractions are consistent, within the errors, for most of the galaxies in the present sample.  {\it Bottom:} The equivalent widths of the lines are close to the intrinsic ratio of about 1.9 (dashed line).     The \ion{Mg}{2} $\lambda$2796 line is  intrinsically stronger in \ion{H}{2} regions, but its ISM optical depth is two times higher than the 2803 \AA\ line.  These observations indicate that, given the competing effects of ISM absorption and intrinsic emission,  absorption is not very significant in setting the doublet ratio in these galaxies. }
\label{mgii_ew_fesc_1} 
\end{figure}

\section{Characterizing \ion{Mg}{2} Escape}  
\label{results} 
The Green Peas in this sample show a range of \ion{Mg}{2} equivalent widths (0 to $9$ \AA) and escape fractions  ($0$ to $0.9$).  Figure   \ref{mgii_ew_fesc_2} 
explores the cause of these variations.    First, the left-hand panel shows that \ion{Mg}{2} equivalent widths correlate with escape fractions.   This trend implies that 
 higher \ion{Mg}{2} equivalent width emission is not necessarily an indication that \ion{Mg}{2} is intrinsically stronger; rather, the ISM is more transparent to \ion{Mg}{2} 
 when the equivalent width is higher.   In fact, the center panel of Figure \ref{mgii_ew_fesc_2} shows that the sources with the strongest \ion{Mg}{2} emission have higher O32 values.    In our Cloudy models (and according to Equations \ref{mg2_fit_1} and \ref{mg2_fit_2}), the higher O32  values are an indication of weaker \ion{Mg}{2}/\oiii\ ratios.  
These lower \ion{Mg}{2}/\oiii\  ratios translate directly to fainter intrinsic \ion{Mg}{2} emission for the sources that have the highest equivalent widths and escape fractions (Figure \ref{mgii_ew_fesc_2}, right).   In summary, the sources with the strongest {\it observed} \ion{Mg}{2} emission are actually those with the weakest {\it intrinsic} emission.

We next look for signs of resonant line radiation transport on the \ion{Mg}{2} lines.  Figure \ref{mgii_ew_fesc_1} presents the \ion{Mg}{2} escape fractions for the 10 Green Peas in this sample, comparing the two lines of the doublet.   The ratio of oscillator strengths describing the ISM absorption on these lines is $f_{2796} /  f_{2803} = 0.61/0.30$, or about a factor of two.   Hence, the optical depths are related: $\tau_{2796} = 2 \times \tau_{2803}$.  However, since these lines are resonant,  the optical depths do not necessarily translate into differing escape fractions.   Rather,   the top panel of Figure \ref{mgii_ew_fesc_1} shows the escape fractions of the lines are mostly consistent within the errors, except for the galaxy with the {\bf  highest dust extinction} and lowest escape fraction in the $\lambda 2796$ line: 1440+4619.  In this case, although we detect \ion{Mg}{2} emission, we also detect absorption on the $\lambda 2796$ line, so the net equivalent width is near zero (before the stellar absorption correction).   Similarly, the bottom panel  of Figure \ref{mgii_ew_fesc_1} shows that the equivalent width ratios,  $W_{2796}/W_{2803}$, are near the intrinsic ratio of about 1.9, inferred from our Cloudy models.     Overall, it appears that the effects of resonant scattering are not significant enough to alter the intrinsic line ratio, or cause the $\lambda$2796 line to have a lower escape fraction.

\begin{figure*} 
\centering
\includegraphics[scale=0.75, viewport=0 0 630 500, clip]{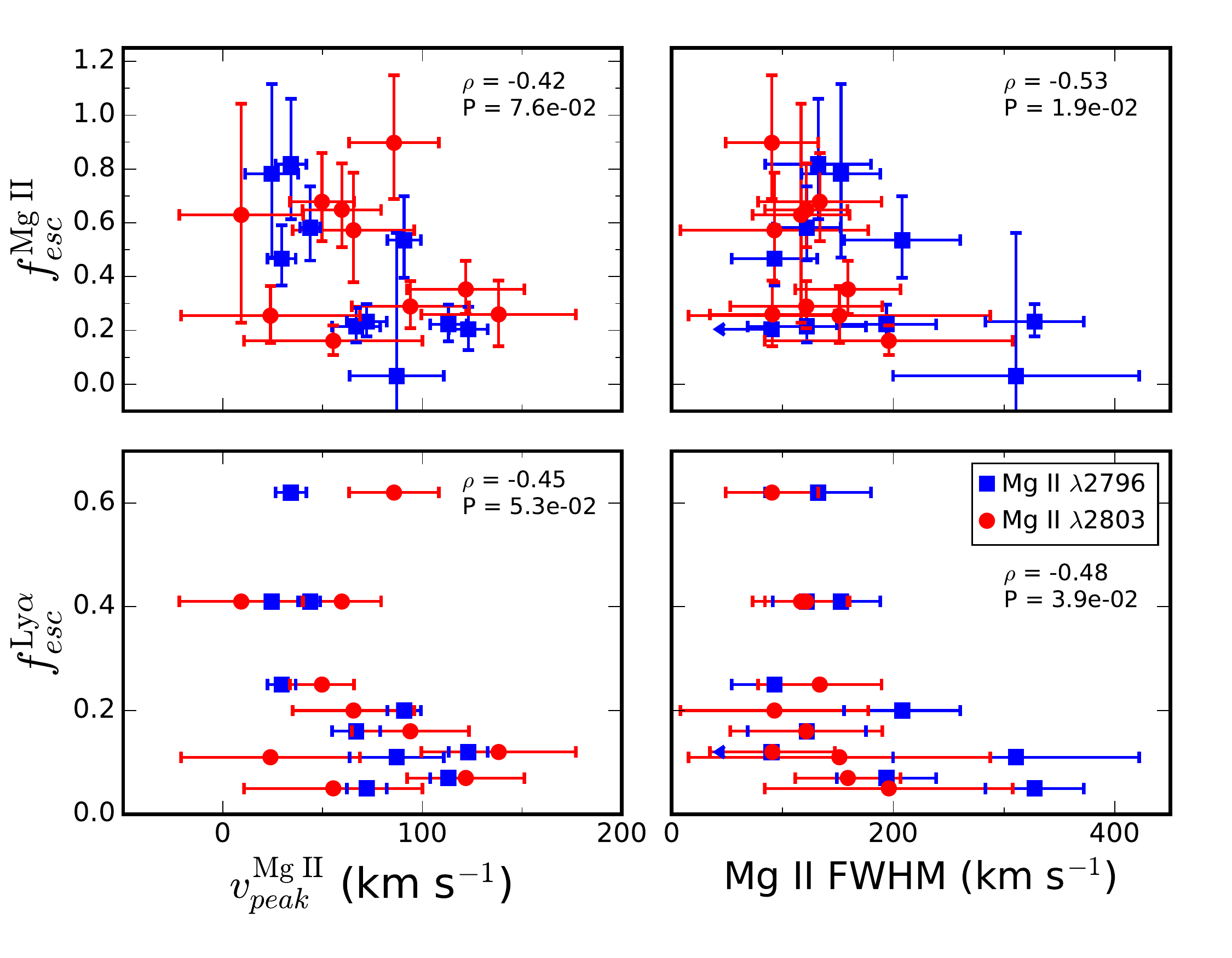} 
\caption{\ion{Mg}{2}  and \lya\ escape fractions show a tentative relation with \ion{Mg}{2} FWHM and peak velocity shift.  These correlations are expected if the escape fractions are determined by low column density of neutral and low-ionization state gas.   When the column densities are low, the redshifting and broadening of the \ion{Mg}{2} lines should be lessened.   The Pearson correlation coefficient, $\rho$, and the probability of the a spurious correlation, $P$, are given in each panel.  The \ion{Mg}{2} FWHM are corrected for instrumental resolution.}
\label{mg2_meas_panel1}  
\end{figure*}

While the \ion{Mg}{2} equivalent width and escape fraction ratios indicate that photon scattering is insufficient to remove $\lambda$2796 photons more than $\lambda 2803$ photons, kinematic measures, in contrast, show evidence of radiation transport. 
As noted in \S \ref{char_mgii}, and shown in Figures \ref{mgii_fig1} and \ref{mgii_fig2}, the \ion{Mg}{2} lines are systematically redshifted by an average of 70 \kms.  Additionally,  in a few cases the \ion{Mg}{2} lines are broader than the optical nebular lines.  While the SDSS spectral lines are only marginally resolved, we have obtained Keck/ESI spectra for the 8/10 Green Peas in the present sample, achieving 75 \kms\ spectral resolution (Henry et al., in prep).   
After correcting for spectral resolution,  we find that the Green Pea 1244+0216 has FWHM of $327 \pm 44$ and $195 \pm 119$ for the $\lambda 2796$ and $\lambda 2803$ lines, respectively, whereas its \hb\ line has a FWHM of only 94 \kms.  Likewise, 1249+1234 has FWHM around $122 \pm 35$ \kms\ in both \ion{Mg}{2} lines, compared to 66 \kms\ in \hb.   Lastly, the $\lambda 2796$ line of 0926+4427 has FWHM of $208\pm 46$, whereas \hb\ is substantially narrower at 106 \kms.   For 1440+4619 and 1442-0209, we do not have a high resolution spectrum for comparison, but  we do note that 1440+4619 has \ion{Mg}{2} lines which are among the broadest in our sample, similar to 1244+0216.  

The broad widths and redshifted emission are analogous to the case of \lya.  Therefore, we may expect the lines to be more redshifted and broadened when the column density of \ion{Mg}{2} is higher, and the escape fractions are lower.  The top panels of Figure \ref{mg2_meas_panel1} test this hypothesis, showing a tentative relation in the expected direction:  higher \ion{Mg}{2} escape fraction may be associated with lower \ion{Mg}{2} peak velocities and smaller FWHM.    The bottom panels of Figure \ref{mg2_meas_panel1} show the same kinematic measures against the \lya\ escape fraction, with a similar tentative correlation.  The Pearson correlation coefficient ($\rho$), and probability of the null hypothesis (P) are shown in each panel,  indicating a 2-8\% likelihood of a spurious correlation.  Additional observations could clarify this relation.  Not only would a broader dynamic range be useful, it is also clear that high-resolution and high signal-to-noise will be needed to measure \ion{Mg}{2} peak velocity shifts of order 100 \kms.  

\begin{figure} 
\includegraphics[scale=0.5, viewport=0 0 630 600, clip]{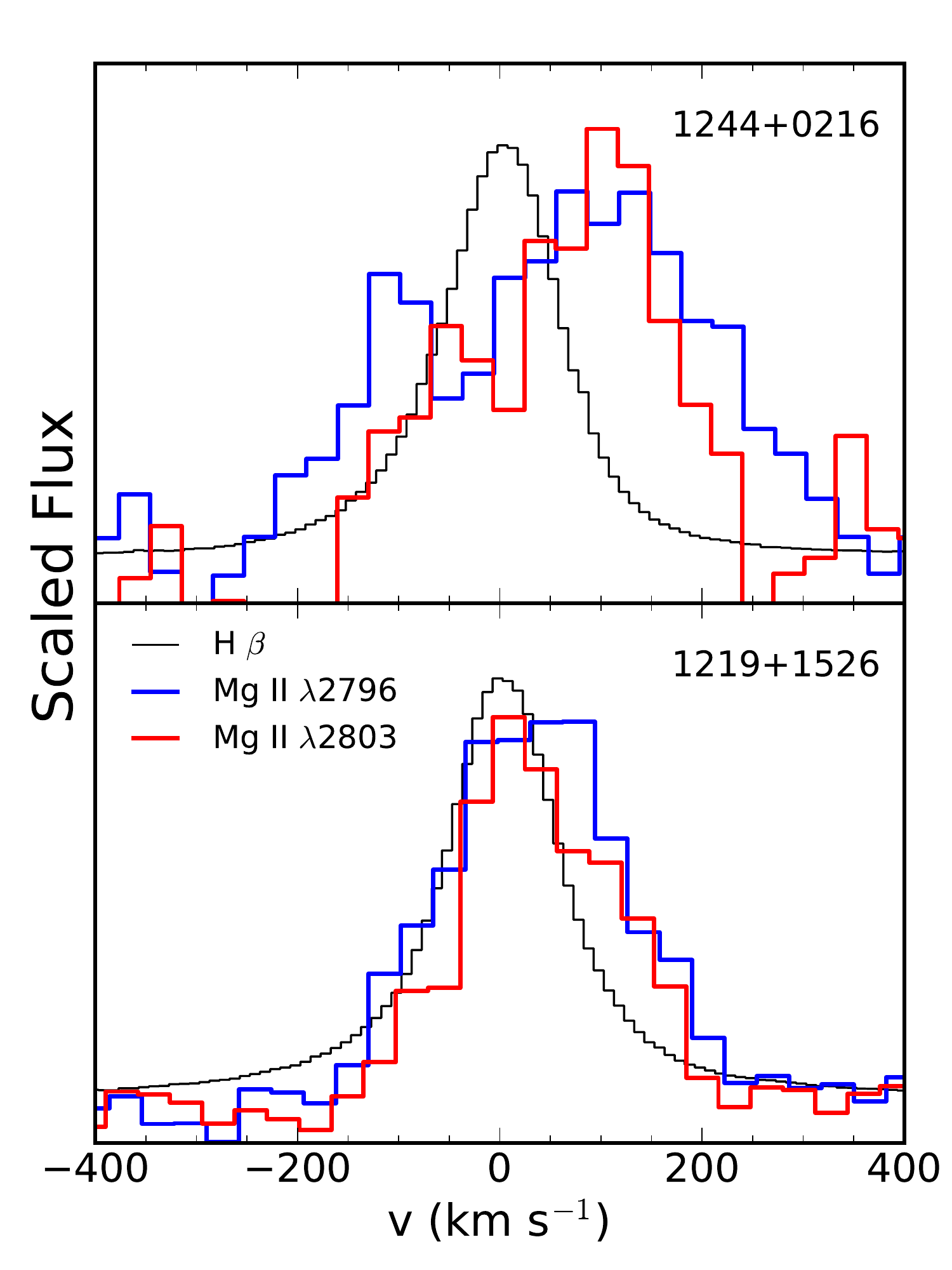}
\caption{\ion{Mg}{2} profiles are compared to \hb\ for two Green Peas.    The cases highlighted here bracket the range of \ion{Mg}{2} and \lya\ escape fractions probed 
by our sample.     Consistent with having higher column densities of low-ionization state gas,  the Green Pea 1244+0216 has double peaked profiles, much broader than H$\beta$.  On the other hand, 1219+1526, which has properties consistent with LyC Leakers \citep{Verhamme17},  shows \ion{Mg}{2} profiles that are mostly symmetric and only marginally broader than \hb\ on the red side.  The line fluxes are scaled arbitrarily to compare the profiles.   }
\label{mg2_prof} 
\end{figure}

A more detailed look at the \ion{Mg}{2} line profiles is shown in Figure \ref{mg2_prof}.   Here we compare the \ion{Mg}{2} and \hb\ lines of two Green Peas, bracketing the range of escape fractions probed by our sample.    This comparison shows clearly that 1244+0216,   which is among the Green Peas with the lowest \lya\ escape fractions, widest \lya\ peak separations, and lowest \ion{Mg}{2} escape fractions, has much broader \ion{Mg}{2} than \hb.  On the other hand, 1219+1526, which has properties similar to Lyman Continuum leakers \citep{Verhamme17},  is only marginally broadened on the red side of the \ion{Mg}{2} lines.   The spectral resolution is comparable between the two observations  (90 \kms\ for the \ion{Mg}{2} spectrum and 75 \kms\ for the \hb),  so the broader \ion{Mg}{2} lines cannot be due to instrumental effects.  We also notice in Figure \ref{mg2_prof} that 1244+0216  has double peaked \ion{Mg}{2} emission on both lines of the doublet.  Here, we are detecting the same effect that shapes the \lya\ profiles of the Green Peas: due to the relatively high optical depth at line center, photons scatter away from the systemic velocity and into the wings of the lines.  Indeed, the somewhat wider peak separation for the $\lambda$2796 line (200 $\pm$ 20 \kms\ vs 170 $\pm$ 40 \kms) is consistent with its higher oscillator strength and optical depth to ISM absorption, relative to the $\lambda$2803 line.     Overall, we conclude that the changes in the peak velocities and FWHM are consistent  with a scenario of varying column densities of neutral and low ionization state gas. 

\begin{figure*} 
\centering
\includegraphics[scale=0.55, viewport=0 0 850 350, clip]{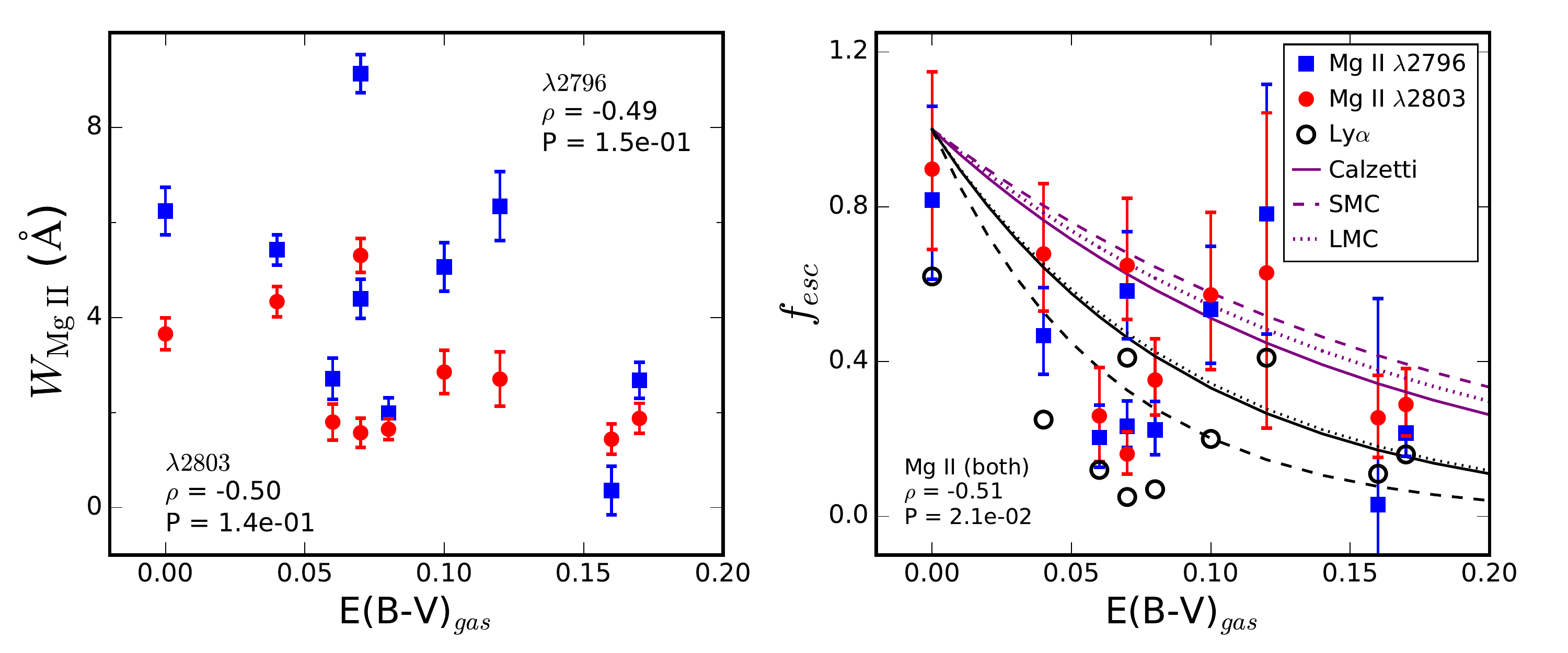} 
\caption{The \ion{Mg}{2} equivalent width and escape fraction are somewhat higher when dust extinction is lower.   The black points show that this trend mirrors the relation seen in \lya.   The Pearson correlation coefficient, $\rho$, and the probability of a spurious correlation are given in each panel, indicating only marginally significant trends.   Extinction curves from \cite{Calzetti}, and the SMC and LMC \citep{Gordon} are  shown in purple for \ion{Mg}{2} and black for \lya.  Despite the tentative correlation,  the measured points show a general trend expected if dust plays a role in setting \ion{Mg}{2} and \lya\ escape.  The fact that the many measurements fall below the extinction curves is an expectation for resonant lines, which are more susceptible to dust extinction than non-resonant lines.     }
\label{mg2_dust}  
\end{figure*}

Next, we explore the role of dust extinction in establishing the \ion{Mg}{2} emission.   Figure \ref{mg2_dust}  shows trends where lower dust extinction is associated with higher \ion{Mg}{2} equivalent width and escape fraction.  Again, the Pearson correlation coefficients are given in each panel, alongside the probability of a spurious correlation.   Although any correlation is only tentative, the trends follow our expectations.  For comparison, we show the starburst dust extinction law from \cite{Calzetti}, along with the Small Magellanic Cloud (SMC), and Large Magellanic Cloud (LMC) from \cite{Gordon}.   While non-resonant lines should follow the appropriate extinction curves, resonant lines are more susceptible to dust attenuation because multiple scatterings increase the likelihood of absorption.   Hence, it is not surprising that many of the \ion{Mg}{2} and \lya\ escape fractions fall below the relevant curves in Figure \ref{mg2_dust}.    While about half the sample lies near the locus where dust extinction alone could explain the \ion{Mg}{2} escape fractions, the other half appears to have too little \ion{Mg}{2} emission for the amount of extinction.   However, we again caution that the \ion{Mg}{2} escape fractions are uncertain in an absolute sense, due to systematics associated with the photoionization modeling and intrinsic \ion{Mg}{2} flux derivation.  Likewise, as we noted in \S \ref{cloudy_sec}, using an SMC, LMC, or Milky Way dust attenuation curve when estimating  intrinsic \ion{Mg}{2} flux (from dust-corrected \oiii) could increase the \ion{Mg}{2} escape fraction by up to 25\% in the sources with the most dust.  Nevertheless, since we do see effects of resonant-line radiation transport on the line profiles, we conclude that the less-than-unity \ion{Mg}{2} escape fractions should probably not be explained by dust extinction without resonant scattering.

\begin{figure*} 
\centering
\includegraphics[scale=0.7, viewport=0 0  700 300, clip]{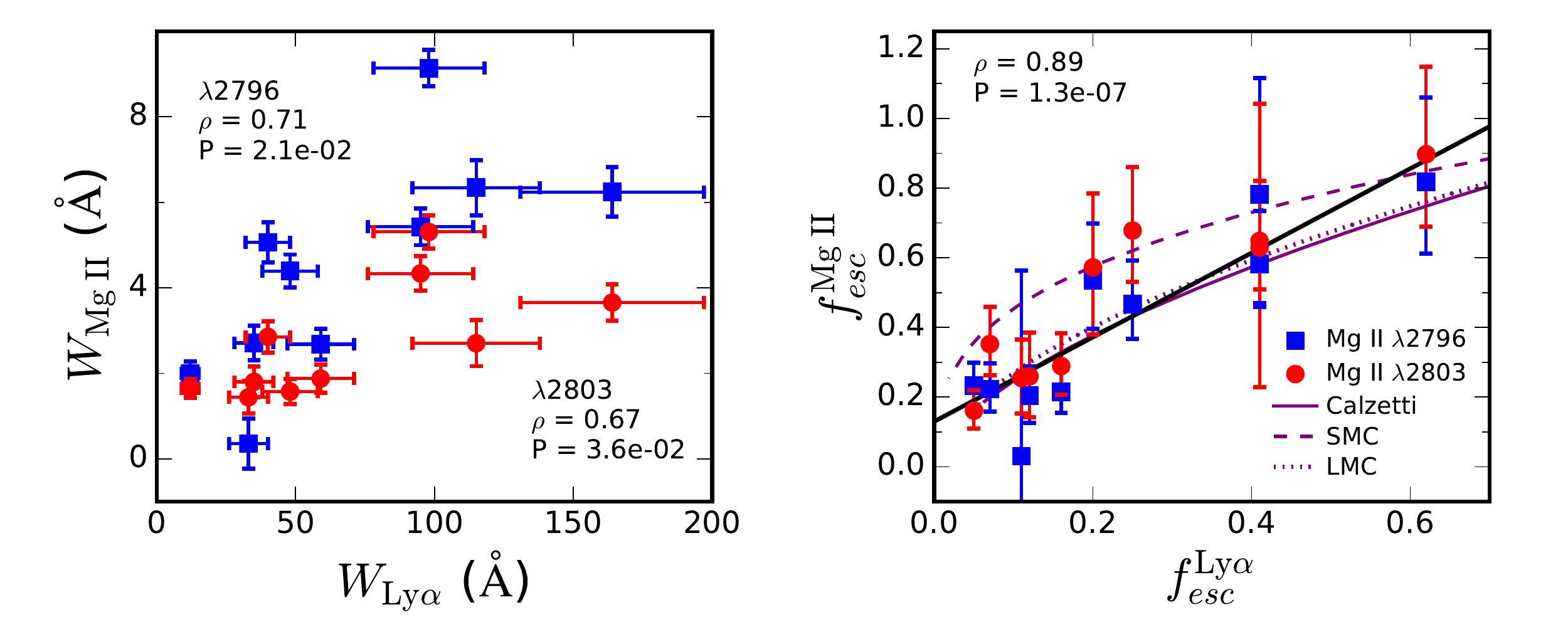} 
\caption{The equivalent widths and escape fractions correlate for \lya\ and \ion{Mg}{2}.   The Pearson correlation coefficient and probability of the null hypothesis are shown in each panel, for the \ion{Mg}{2} lines separately in the equivalent width panel (left), or for combined set of measurements in the escape 
fraction panel (right).   The black line shows a linear fit to the relation, given by Equation \ref{fesc_eq}.   The purple curves show the expectation from dust extinction {\it without} resonant scattering.  We note that  neither the \lya\  or \ion{Mg}{2} escape fractions are corrected for any extended emission that may fall outside the spectroscopic apertures.  }
\label{mg2_meas_panel2}  
\end{figure*}

Since \lya\ and \ion{Mg}{2} are both  shaped by scattering in low-ionization state gas, we might expect their escape fractions to be related.    Indeed, Figure \ref{mg2_meas_panel2} verifies this hypothesis.  The left panel shows that   $W_{\rm Mg\ II}$ correlates with
$W_{\rm Ly\alpha}$, while the right panel shows a tight relation between \fesclya and the \ion{Mg}{2} escape fractions. The Pearson correlation coefficient, and probability of the null hypothesis are listed in each panel.   The relation between the escape fractions-- where we have included both \ion{Mg}{2} lines in the correlation--- is especially robust, with the probability of a spurious correlation around $10^{-7}$.  A linear fit to these data indicates that: 
\begin{equation} 
\label{fesc_eq} 
f_{esc}^{\rm Mg~II} = 0.13 + 1.21~f_{esc}^{\rm Ly\alpha}, 
\end{equation} 
 where both \ion{Mg}{2} lines are included in the fit. 
These correlations are notable:  they confirm that the \lya\ and \ion{Mg}{2} escape are regulated by resonant scattering in the same gas.   Since high \lya\ escape fractions can be a sign of low column densities of \ion{H}{1} gas \citep{Verhamme15}, we infer that high \ion{Mg}{2} escape 
must also be a sign of correspondingly low column densities of \ion{Mg}{2}.   Remarkably, a link between \lya\ and \ion{Mg}{2}  (and the column density of low ionization state gas) opens up the potential for a new diagnostic in the reionization epoch,  where \lya\ and Lyman Continuum are difficult to observe.  We will return to this point in \S \ref{discussion}. 

While a relationship between \ion{Mg}{2} and \lya\  escape fractions is sensible, the observation that these quantities are of the same order is counterintuitive.  The \lya\ and \ion{Mg}{2} transitions have optical depths
to resonant scattering which differ by orders of magnitude.  Hence, one might expect a  substantially lower \fesclya\  for a given \ion{Mg}{2} escape fraction.  Considering dust attenuation highlights
this puzzling  observation.   Figure \ref{mg2_meas_panel2} shows the predicted relationship between the \lya\ and \ion{Mg}{2}, if the lines were non-resonant, for three extinction curves.    All three run through our observations.  Yet, because \lya\ scatters much more than \ion{Mg}{2}, we expect it to be much more impacted by dust.  This effect would place the measurements to the left of the extinction curves.     Although Figure \ref{mg2_dust} suggests that the \ion{Mg}{2} and \lya\ emission see more dust than a non-resonant photon would (for at least some of the sample), contrary to our expectation, Figure \ref{mg2_meas_panel2} suggests that this effect is {\it not} more pronounced for \lya\ than it is for \ion{Mg}{2}. The reason for this observation is unclear.   While similar escape fractions could arise from a scenario where the \lya\ and \ion{Mg}{2} photons escape through pathways with negligible column densities of low-ionization state gas (and dust), we previously showed that Green Peas have uniformly high covering and modest column densities ($N  > 10^{16}$ cm$^{-2}$) in \ion{H}{1} \citep{Henry15}.  Similarly,  \cite{Scarlata09} showed that, for $z\sim 0.3$ \lya\ emitters, there was no need to invoke a scenario where the \lya\ photons escape through dust-free ``holes'' in the ISM.  Therefore, we conclude that more detailed radiation transport modeling would be helpful.  

Even with \ion{Mg}{2} showing showing similarities to \lya, it is surprising  how little it resembles the \ion{Si}{2} $\lambda \lambda 1190, 1193$ profiles that we show in Figures \ref{mgii_fig1} and \ref{mgii_fig2}.   With similar ionization potentials and abundances (assuming a solar relative abundance pattern; \citealt{Grevesse}) the column densities of \ion{Mg}{2} and \ion{Si}{2} should be comparable. However, there are two notable differences between these spectral features.  First,  \ion{Si}{2}  $\lambda \lambda 1190, 1193$  is not produced appreciably in \ion{H}{2} regions.  The Cloudy models presented in \S \ref{cloudy_sec} show that \ion{Si}{2} $\lambda \lambda 1190, 1193$  is 1-3 dex fainter than \ion{Mg}{2}  $\lambda \lambda 2796, 2803$  in almost all cases. Only the 10 Myr aged instantaneous bursts with 
log U $> -2.0$ can have \ion{Si}{2} lines comparable in strength to \ion{Mg}{2}; however, this combination of parameters is probably unphysical, as such high ionization parameters are implausible for models that lack O stars.   So we can conclude that the nebular origin for \ion{Mg}{2} is not matched by \ion{Si}{2}.  The second  difference between \ion{Mg}{2} and \ion{Si}{2} is resonant trapping of the line and continuum photons.  For the \ion{Si}{2} feature, fluorescent fine structure \ion{Si}{2}*  lines at $\lambda$1194 and $\lambda$1197 \AA\ are  coupled with the  \ion{Si}{2} $\lambda \lambda 1190, 1193$  lines.  Photons absorbed in the resonant line can escape through a non-resonant channel,  rather than filling in the absorption in the resonant line. Since \ion{Mg}{2} lacks this non-resonant channel, the absorption is entirely filled by re-emission (as with \lya).   Given these two major differences, we can understand that the line profiles will appear very different in some cases. 
Still, a single, self-consistent model (e.g. \citealt{SP15}) should be able to reproduce the \ion{Mg}{2}, \ion{Si}{2}, and \ion{Si}{2}* when these considerations are taken into account.

\begin{figure*} 
\centering
\includegraphics[scale=0.55, viewport=0 0 850 750, clip]{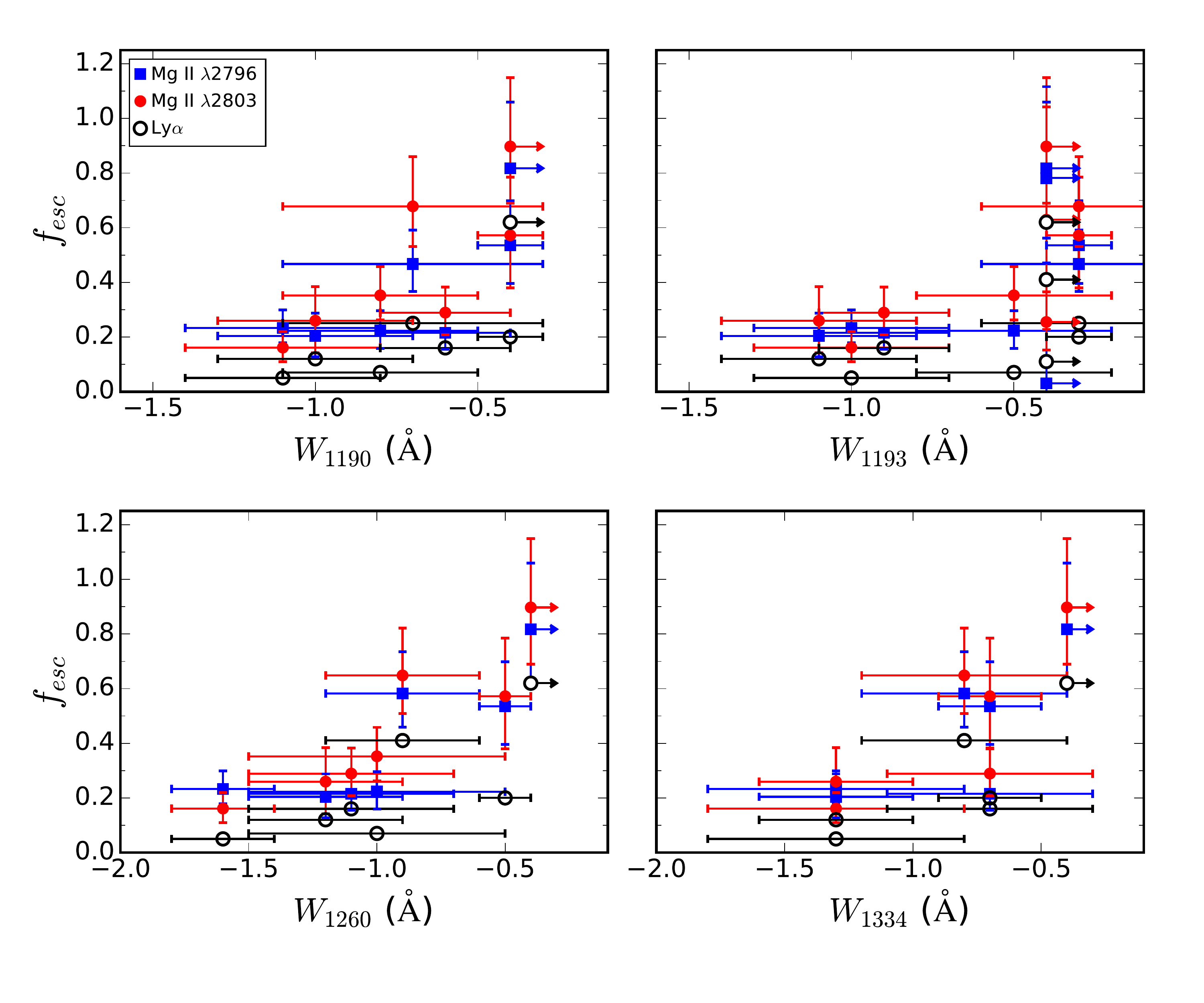} 
\caption{The \ion{Mg}{2} and \lya\ escape fractions are compared to the equivalent widths of the LIS lines of \ion{Si}{2} $\lambda \lambda 1190, 1193$, $\lambda 1260$ and \ion{C}{2} $\lambda 1334$,  measured from the FUV COS spectra.  Previous studies have detected a relation between equivalent widths of these lines and the \lya\ escape fraction or equivalent width \citep{Shapley03, Jones12,  Jones13, Chisholm17}.  While the \ion{Mg}{2} lines are corrected for stellar absorption following \cite{Guseva}, the corrections to the FUV lines are not applied because 
we showed they are  small (see \citealt{Henry15}). }
\label{mg2_lis}  
\end{figure*}

Despite the lack of any qualitative similarities between the \ion{Si}{2} and \ion{Mg}{2} line profiles in Figures \ref{mgii_fig1} and \ref{mgii_fig2}, a quantitative relationship may be expected when we take
the \ion{Mg}{2} escape fractions into account.  Figure \ref{mg2_lis} takes the analogy between \lya\ and \ion{Mg}{2} one step further. 
Equivalent widths in the  low-ionization interstellar (LIS) metal absorption lines have been shown to correlate 
with \lya\ equivalent width or escape fraction \citep{Shapley03, Chisholm17}.
For high redshift  populations,  the interpretation is that the equivalent width of the LIS lines measures the fraction of the 
 background source covered by absorbing ISM \citep{Shapley03, Jones12, Jones13}.  Hence, weaker LIS lines may indicate lower covering fractions, which facilitate higher \lya\ escape fractions and equivalent widths.  On the other hand, \cite{Chisholm17} has argued, using COS spectroscopy of low redshift galaxies, that the same trend (albeit, at a different redshift) can be explained by changing column densities and optically thin LIS lines.  Regardless of the origin, Figure \ref{mg2_lis} tests for a relation by comparing \fesclya\ and $f_{esc}^{\rm Mg\ II}$  to the equivalent widths of four LIS lines: \ion{Si}{2} $\lambda 1190$, $\lambda 1193$, $\lambda 1260$, and \ion{C}{2} $\lambda$1334. This analysis shows that, while the higher escape fraction objects are among those with weaker LIS absorption, any trend is only tentative.   We calculated the Pearson coefficient for a correlation between 
 $f_{esc}^{\rm Mg\ II}$ with the equivalent width of each LIS line,  finding that the probability of the null hypothesis ranges from a few to 30\%.  We note that small samples, covering a small dynamic range cannot always detect a correlation in these quantities \citep{Henry15}.  We also do not detect a correlation between \fesclya\  and the equivalent widths of the LIS lines in Figure \ref{mg2_lis}, even though it has been seen in larger samples (using $W_{Ly\alpha}$ as a proxy for \fesclya; e.g.\ \citealt{Shapley03, Chisholm17}).      Therefore, the large scatter in Figure  \ref{mg2_lis} is not surprising; more observations are needed to determine how $f_{esc}^{\rm Mg\ II}$ varies with the equivalent widths of the LIS lines.

Finally, if \lya\ and \ion{Mg}{2} profiles are shaped by radiation transport in the same scattering gas, we might expect some kinematic measures from the line profiles to correlate.   This idea is supported the fact that our observations were able to resolve double-peaked \ion{Mg}{2} profiles for 1244+0216, which
 has a relatively large  \lya\ peak separation for the Green Peas in this sample.   Since we expect broader, more redshifted emission when the column densities of gas are higher,  we checked for a relation between the \ion{Mg}{2} peak velocities, and the velocities of the red  and blue \lya\ peaks.   Similarly, we compared the FWHM of the \lya\ and \ion{Mg}{2} lines.    In these cases, we see no evidence of a correlation.   
 It is possible, as with the (lack of) correlation between the $f_{esc}^{\rm Mg\ II}$  and the equivalent width of the LIS lines, that a larger sample and more dynamic range is needed.

\section{Summary and Implications} 
\label{discussion}
We have shown that \ion{Mg}{2} appears in emission instead of absorption in Green Pea galaxies.  Previous studies that aim to constrain 
outflows from this spectral feature have noted emission in some cases \citep{Erb12, Martin12}, but the origin of this emission has been unclear. 
 Nevertheless, we showed that Cloudy models predict appreciable emission  from \ion{H}{2} regions (previously noted  by \citealt{Erb12} and \citealt{Jaskot16}) that can easily explain the spectra of these Green Pea galaxies.

Because we have a wealth of other information about the Green Peas, we have characterized their \ion{Mg}{2} emission to 
better understand its use as a diagnostic.  
 Since \ion{Mg}{2} photons are resonantly trapped like \lya, it is not surprising that 
we detect some effects of radiation transport on the lines.   We show that photoionization models can be used to predict the intrinsic \ion{Mg}{2} line fluxes, thereby allowing measurement of \ion{Mg}{2} escape fractions (analogous to \lya).    These escape fractions vary between 0 and 95\%,  and are often lower than the prediction for simple dust extinction without resonant scattering.  We show that lower escape fractions are tentatively associated with \ion{Mg}{2} lines that are broader and more redshifted, which can be an indication of higher column densities of scattering gas.   In particular, the line profiles of the  Green Pea 1244+0216  are remarkable in the sense that they are double peaked and substantially broader than \hb.  This observation makes sense qualitatively, since 1244+0216 is among the objects with the broadest \lya\ peak separations and lowest \ion{Mg}{2} and \lya\ escape fractions.   

At the same time, however, the effects of radiation transport that we observe are substantially weaker than we see for \lya.   First, the line fluxes remain close to their intrinsic ratio.  Likewise, the escape fractions of the $\lambda 2796$ and $\lambda 2803$ are mostly consistent within the errors (except for one case), despite the fact that the blue line has twice the oscillator strength (scattering optical depth) of the red line. Finally, we do not see any evidence of scattered \ion{Mg}{2} emission extending beyond the stellar continuum, in contrast to cases seen by \cite{Rubin11} and \cite{Martin13}.   We acknowledge, however, that this constraint is not particularly strong.  \cite{Yang17b} measure the spatial extent of the \lya\ and UV continuum in 9/10 of the Green Peas in the present sample, and while the \lya\ is 1.3 to 1.9 times more extended than the continuum, the \lya\ sizes are still sub-arcsecond. Since the size of the \ion{Mg}{2} should not exceed \lya, we conclude that our ground-based observations would be unlikely to detect extended \ion{Mg}{2}.  Overall, we conclude that in comparison to \lya, weaker radiation transport effects on \ion{Mg}{2} are sensible, since the column density of \ion{Mg}{2} is orders of magnitude lower than that of \ion{H}{1}.  Still, we are puzzled that resonant scattering in a dusty ISM produces 
\lya\ and \ion{Mg}{2} escape fractions that are of the same order of magnitude.  Since \lya\ photons scatter much more than \ion{Mg}{2} photons, we might expect the \lya\ escape fractions to be substantially lower than \ion{Mg}{2} escape fractions. 

Intriguingly, the \ion{Mg}{2} escape fractions show a tight correlation with the \lya\ escape fraction.   This relation suggests that the same
gas that scatters \lya\ photons and sets their escape is also regulating the \ion{Mg}{2} output. In some ways, this result it not surprising: \ion{Mg}{2} is a low ionization line that serves as a proxy for \ion{H}{1}.    However, in other ways, it is noteworthy.   In spectra of high redshift galaxies, the ISM metal absorption lines have been postulated to trace dense clumps in an outflow \citep{Shapley03}, possibly different from the pervasive \ion{H}{1} component that is indicated by Lyman series absorption lines \citep{Trainor, Reddy}.  If the Green Peas are like high-redshift galaxies, it is not obvious that the dense gas which is most effective at scattering \ion{Mg}{2} photons is co-spatial with large \ion{H}{1} reservoir that scatters \lya\ photons \citep{Henry15}.  
Nevertheless, \ion{Mg}{2} appears to serve as an effective tracer of \ion{H}{1} since the correlation between the escape fractions is surprisingly tight.    More detailed radiation transport modeling will be required to understand the properties of outflows and the ISM/CGM that can produce this relation. 

\subsection{Implications for the Epoch of Reionization} 
The close correspondence between \lya\ and \ion{Mg}{2} could have great utility in the epoch of reionization, where observations of \lya\ are 
difficult or impossible.  Two diagnostics stand out.  First,  \ion{Mg}{2} may be useful for predicting the \lya\ that emerges from the ISM, before it is 
diffused and rendered unobservable by a partially neutral IGM.  Significant efforts have 
been made to measure the neutral hydrogen fraction in the IGM from weakened or absent \lya\ emission at $z>6$ (e.g.,\ \citealt{Mason17}, and references therein).  However, it is unclear how \lya\ would evolve over these redshifts in a completely ionized IGM;  hence, the high-redshift  \lya\ deficit is ill-defined, and inferred IGM neutral fractions are uncertain. Importantly,  \ion{Mg}{2} emission from galaxies would not be impacted by a neutral IGM in the same way.   We can estimate the optical depth for 
scattering \ion{Mg}{2} photons in the IGM at early times, using metallicity measurements from Damped \lya\ Absorber  (DLA) observations.   Observations have indicated  
 [Mg/H] $\sim -2.0$ at $z\sim5$, with  probably declines to higher redshifts \citep{Rafelski14}.  Taking this value as an upper limit, and conservatively assuming that all Mg in a neutral IGM would be in  \ion{Mg}{2}, we estimate that \ion{Mg}{2}/ H  $ \la 10^{-6.4}$, where the solar abundance scale as been adopted \citep{Grevesse}. 
 Since the column density where \lya\ becomes optically thick at line center ($\tau_{Ly\alpha}^{IGM} = 1$)  is around $3\times 10^{13}$ cm$^{-2}$ (the precise quantity depends on the Doppler parameter; \citealt{Verhamme06}), the column density of \ion{Mg}{2} is at most $10^7$ cm$^{-2}$ when \lya\ photons start to see the effects of a neutral IGM.  Hence, the optical depth of \ion{Mg}{2}  at line center will be $\tau_{Mg II}^{IGM} \sim 10^{-6}$,   completely optically thin and unaffected by the neutral IGM. Critically, this means that we could use \ion{Mg}{2} to predict the \lya\ that emerges from the ISM and CGM, before it is impacted by the neutral IGM.   Comparison with observations could then allow more robust measurements of the \lya\ deficit and IGM neutral fraction at very high-redshifts. 

The second reionization-epoch utility for \ion{Mg}{2} is as a diagnostic for Lyman Continuum  (LyC) leakage.   High \lya\ escape fractions have been associated with narrowly spaced \lya\ peaks, low \ion{H}{1} column densities, and direct detection of escaping LyC \citep{Henry15, Verhamme15, Verhamme17, Izotov_nature, Izotov16, Izotov17, Izotov18}.   Because \lya\ and \ion{Mg}{2} escape appear to be governed by radiation transport in the same gas, we postulate that \ion{Mg}{2} may also be used to predict LyC emission.   This hypothesis is supported by the detection of \ion{Mg}{2} in the SDSS spectrum of the LyC emitter, J1154+2443 \citep{Izotov18}, although a higher signal-to-noise spectrum will be required for a detailed comparison to the sources presented here. Ultimately, \ion{Mg}{2} may be particularly useful at $z>6-7$, where \lya\ becomes unobservable due to the  neutral IGM.  With NIRSpec on JWST,  these observations will soon be possible. We can estimate the required exposure time to detect \ion{Mg}{2} from a 26.5th magnitude galaxy with $W = 7$ \AA\ (rest)  on each of the \ion{Mg}{2} lines.  For this calculation, we simulate a 1D spectrum, assuming NIRSpec's G235H disperser with varying levels of noise, and measure the significance of the simulated line.  We  find that a continuum signal-to-noise of 1.5 per pixel is required to detect the \ion{Mg}{2} lines with a signal-to-noise of 5.   Using the JWST Exposure Time Calculator\footnote{https://jwst.etc.stsci.edu/}, we find that 48 hours of observation should meet this goal.  While expensive, 
detections of \ion{Mg}{2} in the reionization epoch should be possible as part of a comprehensive, deep spectroscopic survey.   Clearly, before we can confidently use \ion{Mg}{2} as a diagnostic for reionization,  more work will be needed to characterize its emission and relationship with \lya\ in an ionized IGM.  Nevertheless, the trends that we have presented here appear promising.

\subsection{Implications for the  properties of galaxies and outflows} 
We have shown that the \ion{Mg}{2} feature originates from nebular line photons, which resonantly scatter to escape galaxies, analogous to \lya.
This finding implies that studies which aim to use \ion{Mg}{2} for constraining outflows  {\it must} account for the emitted line photons.  Models which seek to reproduce the \ion{Mg}{2} line profiles with only scattered continuum photons should struggle to reproduce observations (\citealt{Prochaska11, SP15,  Zhu, Bordoloi16}, Carr et al., submitted).  Indeed, in \cite{Erb12}, significant differences between \ion{Mg}{2} and \ion{Fe}{2} line profiles were attributed to a contribution from nebular \ion{Mg}{2} emission.  Our Green Pea observations confirm this interpretation.   In short, when \ion{Mg}{2} is used, measurements of galaxy outflow velocities, column densities, and mass-loss rates will require more sophisticated modeling than has yet been attempted for this spectral feature. Fortunately,  these radiation transport models have already been developed for interpreting \lya\ emission \citep{Dijkstra06, Behrens14, ZW14, Verhamme06, Verhamme08, Verhamme15}. 

The observations that we report  here provide further context for interpreting higher redshift observations. 
Notably, \cite{Finley17} report \ion{Mg}{2} emission is common among low mass galaxies ($M < 10^{9}$ M$_{\sun}$)   at
$z\sim 1$, but less common at high masses.   Within this sample, the lower mass galaxies will have relatively lower metallicities than the high mass galaxies, so they should also have correspondingly higher O32.  Hence, following our photoionization models, we expect that the intrinsic \ion{Mg}{2}/\oiii\ ratio (and possibly the intrinsic \ion{Mg}{2} flux), weakens at lower masses.  Therefore, the strongest \ion{Mg}{2} emitters in \cite{Finley17} probably have 
high \ion{Mg}{2} (and \lya) escape fractions.    These trends among the Green Peas and $z\sim 1$ galaxies could be an indication that a highly ionized ISM (high O32) results in low column densities of neutral and low-ionization state gas, allowing high \ion{Mg}{2} and \lya\ escape fractions. 
Additionally, given our understanding of \ion{Mg}{2} escape, it makes sense that the $z\sim1$ galaxies with  \ion{Mg}{2} emission tend  not to have fluorescent \ion{Fe}{2}* emission \citep{Finley17}. 
While nebular \ion{Mg}{2} emission favors ISM with low column densities of gas, fluorescent \ion{Fe}{2}* is the opposite.  Non-negligible absorption in resonant \ion{Fe}{2}, from higher column densities of gas, is required to produce \ion{Fe}{2}* emission.    The opposite trend-- stronger \ion{Fe}{2}* emission when \ion{Mg}{2} is strong-- reported by \cite{Erb12} is somewhat puzzling.  However, these results are based on composite spectra, and aperture losses on the scattered emission would also tend to correlate the strength of these emission features.

Our analysis of \ion{Mg}{2} also points to an important warning about the use of resonant lines for constraining the physical conditions in \ion{H}{2} regions.  
For example, observations of nebular \ion{C}{4} have been reported from galaxies at both high and low-redshifts \citep{Stark14, Mainali, Schmidt17, Senchyna}, and used to quantify the number of hard ionizing photons from young stars. However,   by analogy with \ion{Mg}{2}, we infer that escape fractions may be anywhere between zero and one.    Indeed, at low impact parameter in the CGM of low-redshift galaxies, \cite{Bordoloi14} find  \ion{C}{4} column densities around $N  = 10^{14}$, which implies \ion{C}{4} optical depth at line center of $\tau \gtrsim 1-2$. Hence, we expect resonant-line radiation transport to be significant; this may explain the apparent redshifting of \ion{C}{4} in some low-redshift galaxies with high-signal to noise UV spectra and known systemic redshifts \citep{Senchyna}.   Although \ion{C}{4} is commonly used as a diagnostic for the physical conditions in HII regions (e.g. \citealt{Stark14, Gutkin, Feltre}) we conclude that this approach is probably unreliable, even when the \ion{C}{4} stellar wind feature can be confidently subtracted.    Despite this complication, the resonant nature of \ion{C}{4} could prove useful for understanding the CGM, similar to \lya\ and \ion{Mg}{2}.   It would be curious to test whether \ion{C}{4} emission and escape\footnote{We acknowledge that photoionization modeling to predict intrinsic \ion{C}{4} may not be as straightforward as with \ion{Mg}{2}.} correlate with \lya\ as \ion{Mg}{2} does.   The \ion{C}{4}  traces different gas than \ion{Mg}{2} and \lya, so we would not necessarily expect the same close relationship that we see between \ion{Mg}{2} and \lya.  Still, any correlation, or lack thereof, could place constraints on the multi-phase nature of the ISM and CGM.

\subsection{Future Work} 
Finally, we conclude that these observations mark an important first step in the 
characterization of \ion{Mg}{2} and its relation with \lya.   We have shown that simultaneous observations of multiple resonance lines probing neutral and low-ionization state gas can give qualitatively consistent results.    We provide evidence that \ion{Mg}{2} may be able to predict the \lya\ that emerges from galaxies, before it is impacted by a neutral IGM at high redshifts.  As a consequence of its close correlation with \lya, \ion{Mg}{2} may also serve as a useful LyC diagnostic at high redshifts.   Overall, larger samples at all redshifts will be useful for strengthening our conclusions about the physical processes that allow \lya, \ion{Mg}{2}, and LyC escape. Indeed, at $z\gtrsim 3$, \ion{Mg}{2} and the full rest-frame optical spectrum can be observed with JWST/NIRSpec, and ground based-data can cover \lya\ at $z<6$.  Hence, the data to test these theories at high redshifts will soon be available.  In the mean time, simultaneous radiation transport modeling of the \lya\ and \ion{Mg}{2} lines will be an intriguing test of models for outflows and the CGM.

\acknowledgements 
The authors thank the anonymous referee for insightful comments, which improved this work.  AH thanks Marc Rafelski,  Nicolas Bouch\'e, Tim Heckman and Molly Peeples for useful suggestions, Anne Jaskot for help with Cloudy models, and James White for reducing  the Keck/ESI spectra.  The authors also
wish to thank the Lorentz Center for facilitating collaboration during the ``Characterizing Galaxies with Spectroscopy with a view for JWST'' workshop.    This work was supported in part by HST GO 13654.

\begin{deluxetable*} {cccccccccc}[!t]
\tablecolumns{10}
\tablecaption{Comparison Data for the Green Peas} 
\tablehead{
\colhead{ID}  &  \colhead{E(B-V)$_{gas}$}  & \colhead{O32}  &\colhead{F([OIII] 5007) }  &   \colhead{12 + log(O/H)}  & \colhead{F(\lya)}  & \colhead{$f_{esc}^{Ly\alpha}$} & \colhead{$W_{Ly \alpha}$}  & $v_{blue}^{peak} $ & $v_{red}^{peak}$    \\
     &     &   &  (10$^{-14}$ erg s$^{-1}$ cm$^{-2}$ ) &   &  (10$^{-14}$ erg s$^{-1}$ cm$^{-2}$ ) &   & (\AA)  & (\kms) &  (\kms)       \\ 
      \colhead{(1)} & \colhead{(2)} & \colhead{(3)} &     \colhead{(4)} & \colhead{(5)} & \colhead{(6)} &  \colhead{(7)} & \colhead{(8)} & \colhead{(9)}  & \colhead{(10) }    
      }
      \startdata
0911+1831 & 0.17 & 0.26  &1.57   &  7.90   & $3.3 \pm 0.1$  &    0.16  & $59 \pm 12$    &  -280   & 90  \\ 
0926+4427 & 0.10 & 0.51  & 4.60   & 7.95  &  $6.0 \pm 0.3$ &    0.20  & $40 \pm 8$       &  -250  & 160 \\ 
1054+5238 & 0.08  & 0.40 &  3.29   &  8.22  & $1.7 \pm 0.2$  &     0.07  & $12 \pm 3$    &  -250  & 160 \\ 
1137+3524  & 0.06 &  0.45 & 5.24    &  8.14 & $3.8 \pm 0.2$   &   0.12   & $ 35 \pm 7$   &  -400 & 150 \\ 
1219+1526 & 0.00  & 1.02  &  5.71    &  7.86 &$13.7 \pm 0.2$  & 0.62   & $164\pm 33$  & -100 & 140 \\ 
1244+0216 &  0.07  & 0.56 &  5.74   &  8.12   & $2.0 \pm 0.1$ &  0.07 & $48\pm 10$       & -280  & 250 \\ 
1249+1234 & 0.07  &  0.55 & 2.52   &   8.10    & $5.4 \pm 0.1$  &  0.41  & $98 \pm 20$    &  \nodata & 70 \\ 
1424+4217  & 0.04 &  0.79 & 7.92    &  7.99     & $8.5 \pm 0.2$ &  0.25 & $95 \pm 19$   & -150 & 230\\ 
1440+4619  & 0.16   & 0.25 &   2.05   & 8.17  & $2.3 \pm 0.2$  & 0.11  & $33 \pm 7$    & -470 & 70  \\ 
1442---0209  & 0.12  &  0.83 &   2.05  &   7.93  & $5.1 \pm 0.2$  &  0.41   &$115 \pm 23$  & -230  & 100   
\enddata
\label{lya_table} 
\tablecomments{Comparison data for the Green Peas are drawn from SDSS and HST/COS spectroscopy.  \lya\ measurements are either taken from \cite{Henry15}, or, in the case of 1440+4619 and 1442-0209 measured in the same way.   Column descriptions: (1) object ID; (2) nebular dust extinction, calculated from \ha/\hb, using the \cite{Calzetti} extinction curve; (3) O32 $\equiv$ log(\oiii\ $\lambda$5007 / \oii\ $\lambda$3727),corrected for dust; (4) \oiii\ $\lambda$5007 flux, corrected for dust; (5) gas-phase metallicity, calculated following \cite{Berg15}.  The values given here differ slightly from those in \cite{Henry15}, where they were instead taken from \cite{Izotov11};   (6) The \lya\ flux measured from the COS spectra, corrected for Milky Way foreground attenuation using \citealt{Schlafly} extinction
 measurements and the \citealt{fm_unred} extinction law; (7) the \lya\ escape fraction; (8) the rest-frame \lya\ equivalent width; (9) the velocity marking the blue peak of the \lya\ line profile; (10) the velocity marking the red peak of the \lya\ line profile. } 
\end{deluxetable*}

\clearpage 
\begin{turnpage} 
\begin{deluxetable*} {crcccccccccccc}[!t]
\tablecolumns{14}
\tablecaption{\ion{Mg}{2} Measurements from Green Peas} 
\tablehead{
\colhead{ID}  &  \colhead{$W_{2796}$}  & \colhead{$W_{2803}$} &   & \colhead{F$_{2796}$} &\colhead{F$_{2803}$}   &   & $v_{2796} $ & $v_{2803}$ & \colhead{FWHM$_{2796}$} & \colhead{FWHM$_{2803}$}  &   & \colhead{$f_{esc}$ (2796)}  &  \colhead{$f_{esc}$ (2803)}  \\ 
\\ 
\cline{2-3}  \cline{5-6}   \cline{8-11} \cline{13-14}  \\ 
  &     \multicolumn{2}{c}{(\AA)}  & &   \multicolumn{2}{c}{($10^{-16}$ erg s$^{-1}$ cm$^{-2}$)} & &  \multicolumn{4}{c}{(\kms)}  & &   &  \\ 
 \colhead{(1)} & \colhead{(2)} & \colhead{(3)} &   &    \colhead{(4)} & \colhead{(5)} & & \colhead{(6)} &   \colhead{(7)} & \colhead{(8)} & \colhead{(9)} &  & \colhead{(10)}& \colhead{(11)}
} 
\startdata
0911+1831 & $2.9 \pm 0.4$ & $1.9  \pm 0.3$  &   &    $5.5 \pm 1.0$ & $3.7 \pm 0.8$  &  &  $67 \pm 12$  & $94  \pm 28$ & $122 \pm 59$   & $121 \pm 68$ & &$0.21^{+0.07}_{-0.06}$ & $0.29^{+0.08}_{-0.08}$  \\ 
\rule{0pt}{3ex}%
0926+4427 & $5.1  \pm 0.5$ & $2.9 \pm 0.5$ &    &    $21.7 \pm 3.4$  & $12.2 \pm 3.3$  & & $91 \pm 10$  & $65  \pm 32$ & $208 \pm 46$  & $93 \pm 69$   & & $0.53^{+0.16}_{-0.13}$   & $0.57^{+0.21}_{-0.19}$  \\ 
\rule{0pt}{3ex}%
1054+5238 & $3.0  \pm 0.3$ &  $1.6 \pm 0.2$  &  &     $8.4 \pm 1.7 $ & $6.9 \pm 1.1$  & &  $113 \pm 9$  & $122  \pm 22$ & $193 \pm 38$ & $159 \pm 46$ &  & $0.23^{+0.07}_{-0.05}$ &  $0.36^{+0.11}_{-0.09}$   \\   
\rule{0pt}{3ex}%
1137+3524  & $2.7 \pm 0.4$ & $1.8\pm 0.4$  & &    $10.8 \pm 4.1$  & $7.2  \pm 2.9$  & &  $123 \pm 9$ & $138 \pm 44$  & $<90$  & $91 \pm 59 $   &  & $0.20^{+0.07}_{-0.07}$ &  $0.26^{+0.11}_{-0.08}$ \\
\rule{0pt}{3ex}%
1219+1526 & $6.2\pm 0.5$  & $3.7 \pm0.3$  &  &    $14.1 \pm 2.0$   & $8.6 \pm 0.9$ &   & $34  \pm 7$& $86 \pm 17$  & $132 \pm 45$ &$90 \pm 40$   &  & $0.80^{+0.24}_{-0.21}$  & $0.90^{+0.25}_{-0.22}$   \\  
\rule{0pt}{3ex}%
1244+0216 & $4.4 \pm 0.4$ &  $1.6 \pm 0.3$  &   &   $10.6  \pm 1.2 $  & $3.8 \pm 0.9$  & &  $72\pm 8$  & $55  \pm 38$ & $327 \pm 44$  & $195 \pm 119$   & &$0.24^{+0.07}_{-0.04}$ & $0.16^{+0.06}_{-0.05}$ \\
\rule{0pt}{3ex}%
1249+1234 & $9.1  \pm 0.4$ & $5.3 \pm 0.4$   & &   $12.0  \pm 0.5$   & $7.1 \pm 0.5$   &  &  $ 44\pm 4$& $60 \pm 23$  & $122 \pm 31$  &  $121 \pm 38$  & & $0.58^{+0.16}_{-0.13}$ & $0.65^{+0.17}_{-0.13}$  \\
\rule{0pt}{3ex}%
1424+4217  & $5.4 \pm 0.3$  & $4.3 \pm 0.3$  & &    $18.1  \pm 1.1$ & $14.1 \pm 0.9$  &   & $29 \pm 5$ & $50 \pm 13$  & $92 \pm 41$   & $133 \pm 52$  & &$0.46^{+0.13}_{-0.10}$ & $0.68^{+0.18}_{-0.15}$ \\ 
\rule{0pt}{3ex}%
1440+4619  &  $0.4 \pm 0.5$  &  $1.4  \pm 0.3$ &  & $1.0^{+17.2}_{-1.0}$\tablenotemark{a} &   $4.3  \pm 1.4$ & & $87 \pm 21$ & $24  \pm 42$ & $311 \pm 110$ & $151 \pm 148$  & & $0.03^{+0.54}_{-0.03}$ & $0.25^{+0.11}_{-0.11}$  \\  
\rule{0pt}{3ex}%
1442--0209  & $6.3 \pm  0.7$  & $2.7\pm 0.6$  &  & $7.2 \pm 2.5$  & $3.1 \pm 2.0$  & &  $24\pm 14$ & $9 \pm 25$    & $153 \pm 37$   & $116  \pm 44$   & & $0.78^{+0.34}_{-0.30}$ & $0.61^{+0.42}_{-0.40}$  
\enddata
\label{mgii_table} 
\tablecomments{Measurements of the \ion{Mg}{2} emission lines from our MMT spectra.  Column descriptions: (1) object ID; (2-3) rest frame equivalent widths for the $\lambda 2796$ and $\lambda 2803$ lines.   A stellar absorption correction of 0.3 \AA\ is added. The $\lambda$2796 line is measured including any blueshifted absorption, hence, 1440+4619 has $W_{2796} \sim 0$ \AA\ before the stellar absorption correction.   (3-4) line fluxes are corrected for stellar absorption,  slit-losses (see \S \ref{data}),  and the Milky Way foreground attenuation.  The latter correction uses the \cite{Schlafly} extinction measurements and the \cite{fm_unred} extinction law.   (5-6) velocities marking the peaks of the lines, (7-8) FWHM of the lines, corrected for instrumental resolution (9-10) \ion{Mg}{2} escape fractions, calculated by correcting for stellar absorption, slit losses and Milky Way extinction, but not  corrected for internal dust extinction. Errors on the \ion{Mg}{2} escape fraction include the measurement uncertainty on the observed flux and a 0.1 dex uncertainty on the predicted intrinsic \ion{Mg}{2} flux.}
\tablenotetext{a}{The measured equivalent width on this line is small compared to the stellar absorption correction, so the flux is multiplied by a factor of 6.2.   Prior to this correction, we measure F$_{2796} = 0.16^{+2.8}_{-0.16}$ $\times 10^{-16}$ erg s$^{-1}$ cm$^{-2}$. }
\end{deluxetable*}
\clearpage 
\end{turnpage}



\begin{thebibliography} 


\bibitem[Angel et al.(1979)]{bluechannel} 
Angel, J.~R.~P., Hilliard, R.~L., \& Weymann, R.~J.\ 1979, The MMT and the Future of Ground-Based Astronomy, 385, 87 


\bibitem[Behrens et al.(2014)]{Behrens14} 
Behrens, C., Dijkstra, M., Neimeyer, J.~C. 2014, \aap, 563, 77 

\bibitem[Berg et al.(2015)]{Berg15}
Berg, D.~A.,  Skillman, E.~D., Croxall, K.~V., et al. 2015, \apj, 806, 16 

\bibitem[Boeshaar et al.(1982)]{Boeshaar} 
Boeshaar, G.~O., Harvel, C.~A., Mallama, A.~D., et al.\ 1982, NASA Conference Publication, 2238, 374

\bibitem[Bordoloi et al.(2014)]{Bordoloi14}
Bordoloi, R., Tumlinson, J., Werk, J., et al. 2014, \apj, 796, 136 

\bibitem[Bordoloi et al.(2016)]{Bordoloi16} 
Bordoloi, R., Rigby, J.~R., Tumlinson, J., et al. 2016, \mnras, 458, 1891 

\bibitem[Calzetti et al.(2000)]{Calzetti} 
Calzetti, D. Armus, L., Bohlin, R.~C., et al. 2000, \apj, 533, 682 

\bibitem[Cardamone et al.(2009)]{Cardamone} 
Cardamone, C., Schawinski, K., Sarzi, M., et al. 2009, \mnras, 399, 1191

\bibitem[Cardelli et al.(1989)]{ccm} 
Cardelli, J.~A., Clayton, G.~C., Mathis, J.~S. 1989, \apj, 345, 245

\bibitem[Chisholm et al.(2017)]{Chisholm17}
Chisholm, J., Orlitov\'a, Schaerer, D., et al. 2017, \aap, 605, 67 

\bibitem[Coil et al.(2011)]{Coil} 
Coil, A.~L., Weiner, B.~J., Holz, D.~E., et al. 2011, \apj, 743, 46

\bibitem[Davies et al.(2017)]{Davies17} 
Davies, B., Kudritzki, R.-P., Lardo, C., et al., arXiv:1708.08948 

\bibitem[De Cia et al.(2016)]{deCia} 
De Cia,A., Ledoux, C., Mattson, L., et al. 2016, \aap, 596, 97

\bibitem[Dijkstra et al.(2006)]{Dijkstra06}
Dijkstra, M. Haiman, Z., \& Spaans, M. 2006, \apj, 649, 37 
\bibitem[Dopita \& Sutherland(2003)]{DS03} Dopita, M.~A., \& Sutherland, R.~S.\ 2003, Astrophysics of the diffuse universe, Berlin, New York: Springer, 2003.~Astronomy and astrophysics library, ISBN 3540433627,  

\bibitem[Erb et al.(2010)]{Erb10}
Erb, D.~K., Pettini, M., Shapley, A.~E., et al. 2010, \apj, 719, 1168 

\bibitem[Erb et al.(2012)]{Erb12} 
Erb, D.~K., Quider, A.~M., Henry, A.~L., \& Martin, C.~L. 2012, \apj, 759, 26 

\bibitem[Eldridge \& Stanway(2009)]{bpass} 
Eldridge, J.~J., \& Stanway, E.~R. 2009, \mnras, 419, 479 

\bibitem[Eldridge et al.(2017)]{bpass2p1}
Eldridge, J.~J., Stanway, E.~R., Xiao, L., et al. 2017, PASA, 34, e058 

\bibitem[Feltre et al.(2016)]{Feltre}
Feltre, A., Charlot, S., Gutkin, J. 2016, \mnras, 456, 3354 

\bibitem[Ferland et al.(2017)]{cloudy17} 
Ferland, G.~J., Chatzikos, M.,  Guzm\'an, F., et al.,  arXiv:1705.10877

\bibitem[Filippenko(1982)]{Fillippenko}
 Filippenko, A.~V.\ 1982, \pasp, 94, 715 

\bibitem[Finley et al.(2017)]{Finley17}
Finley, H., Bouch\'e, N., Contini, T., et al. 2017, \aap, in press, arXiv:1710.09195

\bibitem[Finkelstein et al.(2007)]{Finkelstein07}
Finkelstein, S.~L, Rhoads, J.~E., Malhotra, S., Pirzkal, N., \& Wang, J.~X. 2007, \apj, 660, 1023 

\bibitem[Fitzpatrick(1999)]{fm_unred} 
Fitzpatrick, E.~L. 1999, \pasp, 111, 63 

\bibitem[Gordon et al.(2003)]{Gordon}
Gordon, K.~D., Clayton, G.~C., Misselt, K.~A., Landolt, A.~U., \& Wolff, M.~J. 2003, \apj, 594, 279

\bibitem[Garnett(1992)]{Garnett92}
Garnett, D.~R., 1992, \aj, 103, 1330 

\bibitem[Gawiser et al.(2007)]{Gawiser07}
Gawiser, E.~J., Francke, H., Lai, K., et al. 2007, \apj, 671, 278 


\bibitem[Gehrels(1986)]{Gehrels} 
Gehrels, N. 1986, \apj, 303, 336 

\bibitem[Grevesse et al.(2010)]{Grevesse} 
Grevesse, N., Asplund, M., Sauval, A.~J., \& Scott, P. 2010, Ap \& SS, 328, 179 

\bibitem[Guseva et al.(2013)]{Guseva} 
Guseva, N.~G., Izotov, Y.~I., Fricke, K.~J.,  \& Henkel, C. 2013, \aap, 555, 90 

\bibitem[Gutkin et al.(2016)]{Gutkin}
Gutkin, J., Charot, S., Bruzual, G. 2016, \mnras, 462, 1757

\bibitem[Hanuschik(2003)]{uves} 
 Hanuschik, R.~W.\ 2003, \aap, 407, 1157 


\bibitem[Hayes et al.(2014)]{Hayes14} 
Hayes, M., \"Ostlin, G., Duval, F., et al. 2014, \apj, 782, 6 

\bibitem[Henry et al.(2015)]{Henry15} 
Henry, A., Scarlata, C., Erb, D., \& Martin, C. 2015, \apj, 809, 19 

\bibitem[Izotov et al.(2011)]{Izotov11} 
Izotov, Y.~I., Guseva, N.~G., Thuan, T.~X. 2011,  \apj, 728, 161 

\bibitem[Izotov et al.(2016a)]{Izotov_nature} 
Izotov, Y., Orlitov\'a, I., Schaerer, D., et al. 2016a, Nature, 529, 178

\bibitem[Izotov et al.(2016b)]{Izotov16}
Izotov, Y., Schaerer, D., Thuan, T.~X., et al. 2016b, \mnras, 461, 3683

\bibitem[Izotov et al.(2017)]{Izotov17}
Izotov, Y, Schaerer, D., Worseck, G., et al. 2017, arXiv:1711.11449 

\bibitem[Izotov et al.(2018)]{Izotov18} 
Izotov, Y., Schaerer, D., Worseck, G. 2018, \mnras, 474, 4514 

\bibitem[Jaskot \& Ravindranath(2016)]{Jaskot16} 
Jaskot, A.~E., \& Ravindranath, S. 2016, \apj, 833, 136 

\bibitem[Jones et al.(2012)]{Jones12} 
Jones, T., Stark, D.~P.,  Ellis, R. 2012, \apj, 751, 51 

\bibitem[Jones et al.(2013)]{Jones13}
Jones, T.~A., Ellis, R.~S., Schenker, M.~A., Stark, D.~P. 2013, \apj, 779, 52 

\bibitem[Karman et al.(2016)]{Karman}
Karman, W., Grillo, C., Balestra, I. 2016, \aap, 585, 27 

\bibitem[Kornei et al.(2012)]{Kornei12} 
Kornei, K.~A., Shapley, A.~E., Martin, C.~L., et al. 2012, \apj, 758, 135 

\bibitem[Kornei et al.(2013)]{Kornei13}
Kornei, K.~A., Shapley, A.~E., Martin, C.~L., et al. 2013, \apj, 774, 50 

\bibitem[Kroupa(2001)]{Kroupa} 
Kroupa, P. 2001, \mnras, 322, 231 

\bibitem[Leclercq et al.(2017)]{Leclercq} 
Leclercq, F., Bacon, R., Wistotzki, L. et al. 2017, \aap, in press, arXiv:1710.10271

\bibitem[Leitherer et al.(1999)]{Leitherer99}
Leitherer, C., Schaerer, D., Goldader, J., et al. 1999, \apjs, 123, 3 

\bibitem[Leitherer et al.(2014)]{Leitherer14} 
Leitherer, C., Ekstr\"om, S., Meynet, G., et al. 2014, \apjs,212, 14  


\bibitem[Mainali et al.(2017)]{Mainali} 
Mainali, R., Kollmeier, J.~A., Stark, D.~P., et al. 2017, \apj, 836, 14L

\bibitem[Martin \& Bouch\'e(2009)]{MB09}
Martin, C.~L. \& Bouch\'e, N. 2009, \apj, 703, 1394

\bibitem[Martin et al.(2012)]{Martin12} 
Martin, C.~L., Shapley, A.~E., Coil, A.~L., et al. 2012, \apj, 760, 127 

\bibitem[Martin et al.(2013)]{Martin13} 
Martin, C.~L., Shapley, A.~E.,  Coil, A.~L., et al. 2013, \apj, 770, 41 

\bibitem[Mason et al.(2017)]{Mason17} 
Mason, C., Treu, T., Dijkstra, M. et al. 2017, arXiv:1709.5356 

\bibitem[Moustakas \& Kennicutt(2006)]{MK06} 
Moustakas, J., \& Kennicutt, R.~C. 2006, \apj, 164, 81 


\bibitem[Prochaska et al.(2011)]{Prochaska11} 
Prochaska, J.~X., Kasen, D., \& Rubin, K. 2011, \apj, 734, 24 

\bibitem[Rafelski et al.(2014)]{Rafelski14}
Rafelski, M., Neeleman, M., Fumagalli, M., Wolfe, A.~M., \& Prochaska, J.~X. 2014, \apj, 782, 29L

\bibitem[Reddy et al.(2016)]{Reddy} 
Reddy, N.~A., Steidel, C.~C., Pettini, M., Bogosavljevi\'c, M., Shapley, A. 2016, \apj, 828, 108 


\bibitem[Rigby et al.(2014)]{Rigby14} 
Rigby, J.~R., Bayliss, M.~B., Gladders, M.~D., et al. 2014, \apj, 790, 44 

\bibitem[Rubin et al.(2010)]{Rubin10} 
Rubin, K.~H.~R., Weiner, B.~J., Koo, D.~C., et al. 2010, \apj, 719, 1503 

\bibitem[Rubin et al.(2011)]{Rubin11}
Rubin, K.~H.~R., Prochaska, J.~X., M\'enard, B., et al. 2011, \apj, 728, 55 


\bibitem[Scarlata et al.(2009)]{Scarlata09}
Scarlata, C. Colbert, J., Teplitz, H.~I., et al. 2009, \apj, 704, 98L

\bibitem[Scarlata \& Panagia(2015)]{SP15} 
Scarlata, C. \& Panagia, N. 2015, \apj, 801, 43 

\bibitem[Schlafly \& Finkbeiner(2011)]{Schlafly} 
Schlafly, E.~F., \& Finkbeiner, D.~P. 2011, \apj, 737, 103  

\bibitem[Schmidt et al.(2017)]{Schmidt17}
Schmidt, K.~B., Huang, K.-H., Treu, T., et al. 2017, \apj, 839, 17 

\bibitem[Senchyna et al.(2017)]{Senchyna}
Senchyna, P., Stark, D.~P., Vidal-Garc\'ia, A., et al. 2017, \mnras, 472, 2608


\bibitem[Shapley et al.(2003)]{Shapley03} 
Shapley, A.~E. Steidel, C.~C., Pettini, M., \& Adelberger, K.~L. 2003, \apj, 588, 65 


\bibitem[Stark et al.(2014)]{Stark14} 
Stark, D.~P., Richard, J.~P., Siana, B., et al. 2014, \mnras,445, 3200 

\bibitem[Steidel et al.(2016)]{Steidel16} 
Steidel, C.~C., Strom, A.~L., Pettini, M., et al. 2016, \apj, 826, 159 

\bibitem[Tang et al.(2014)]{Tang} 
Tang, Y., Giavaoislco, M., Guo, Y., \& Kurk, J. 2014, \apj, 793, 92 

\bibitem[Trainor et al.(2016)]{Trainor} 
Trainor, R.~F. Strom, A.~L., Steidel, C.~C., Rudie, G.~C. \apj, 2016, 832, 171 


\bibitem[Verhamme et al.(2006)]{Verhamme06} 
Verhamme, A.,  Schaerer, D., \& Maselli, A., \aap, 460, 397 

\bibitem[Verhamme et al.(2008)]{Verhamme08} 
Verhamme, A., Schaerer, D., Atek, J., Tapken, C. 2008, \aap, 491, 89 

\bibitem[Verhamme et al.(2015)]{Verhamme15}
Verhamme, A., Orlitov\'a, I., Schaerer, D., \& Hayes, M. 2015, \aap, 578, 7 

\bibitem[Verhamme et al.(2017)]{Verhamme17} 
Verhamme, A., Orlitov\'a, I., Schaerer, D., et al. 2017, \aap, 597, 13 

\bibitem[Weiner et al.(2009)]{Weiner09}
Weiner, B.~J., Coil, A.~L., Prochaska, J.~X., et al. 2009, \apj, 692, 187 

\bibitem[Wisotzki et al.(2016)]{Wisotzki} 
Wisotzki, L., Bacon,, R., Blaizot, J. et al. 2016, A\&A, 587, 98 

\bibitem[Yang et al.(2017a)]{Yang17} 
Yang, H.,  Malhotra, S.,  Gronke, M., et al.\, 2017a, \apj, 844, 171  

\bibitem[Yang et al.(2017b)]{Yang17b}
Yang, H., Malhotra, S., Rhoads, J., et al. 2017b, \apj, 838, 4

\bibitem[Zahid et al.(2017)]{Zahid17} 
Zahid, H.~J.,  Kudritzki, R.~P., Conroy, C. Andrews, B., Ho, I.-T. 2017, arXiv:1708.07107 

\bibitem[Zheng \& Wallace(2014)]{ZW14} 
Zheng, Z., \& Wallace, J. 2014, \apj, 794, 116 

\bibitem[Zhu et al.(2015)]{Zhu} 
Zhu, G.~B., Comparat, J., Keib, J.-P., et al. 2015, \apj, 815, 48 




\end{thebibliography}
\end{document}